\documentclass[reprint,amsmath,amssymb,aps,superscriptaddress,nofootinbib]{revtex4-1}
\usepackage[utf8]{inputenc}
\usepackage{hyperref}
\hypersetup{
    colorlinks=true,
    linkcolor=blue,
    filecolor=magenta,      
    urlcolor=cyan,
    pdfpagemode=FullScreen,
}
\usepackage{verbatim}
\usepackage{upgreek}

\usepackage{graphicx}
\usepackage[dvipsnames]{xcolor}
\usepackage{soul,xcolor}
\usepackage{xcolor, ulem, color, framed}

\newcommand{\bt}{\mathbf{b}}
\newcommand{\Pt}{\mathbf{P}}
\newcommand{\kt}{\mathbf{k}}

\begin{document}

\title{Exponential Approach to the Hydrodynamic Attractor\\in Yang-Mills Kinetic Theory}

\author{Xiaojian Du} 
\email{xjdu@physik.uni-bielefeld.de}
\affiliation{Fakult\"at f\"ur Physik, Universit\"at Bielefeld, D-33615 Bielefeld, Germany}

\author{Michal P.\ Heller} 
\email{michal.p.heller@ugent.be}
\affiliation{Department of Physics and Astronomy, Ghent University, 9000 Ghent, Belgium}

\author{Sören Schlichting }
\email{sschlichting@physik.uni-bielefeld.de}
\affiliation{Fakult\"at f\"ur Physik, Universit\"at Bielefeld, D-33615 Bielefeld, Germany}

\author{Viktor Svensson}
\email{viktor.svensson@ftf.lth.se}
\affiliation{Division of Solid State Physics and NanoLund, Lund University, S-221 00 Lund, Sweden}

\date{\today}

\begin{abstract}
    We use principal component analysis to study the hydrodynamic attractor in Yang-Mills kinetic theory undergoing the Bjorken expansion with Color Glass Condensate initial conditions. The late time hydrodynamic attractor is characterized by a single principal component determining the overall energy scale. How it is reached is governed by the disappearance of single subleading principal component characterizing deviations of the pressure anisotropy, the screening mass and the scattering rate. We find that for wide range of couplings the approach to the hydrodynamic attractor at late times is well described by an exponential. Its decay rate dependence on the coupling turns out to translate into a simple dependence on the shear viscosity to entropy density ratio.
\end{abstract}

\maketitle
%\tableofcontents
\section{Introduction}

Ultrarelativistic heavy-ion collisions at RHIC and LHC produce a collective state of matter comprised of the fundamental quark and gluon constituents of strong interactions. Describing ab initio formation and evolution of this quark-gluon plasma (QGP) presents an important theoretical problem. Since a first principles description directly using the theory of strong force, quantum chromodynamics (QCD), remains an outstanding challenge, the standard model of the space-time evolution of heavy-ion collisions is based on effective descriptions of QCD~\cite{Heinz:2013wva}. The initial non-equilibrium dynamics after the collision has been addressed in a variety of different microscopic models~\cite{schlichting2019first, Berges:2020thermalization}. Eventually, on a time scale of $\sim 1\,\rm{fm}/c$, the QGP can be well described by relativistic hydrodynamics. This is to some extent surprising, as the system can feature large spatial and temporal gradients and remains significantly out of equilibrium for a much longer period of time, as indicated e.g. by large values of the pressure anisotropy.

Due to these concerns the applicability of relativistic hydrodynamics in high-energy collisions of heavy and light nuclei has repeatedly been questioned~\cite{Romatschke:2016hle}, and different proposals have been put forward to extend the applicability of hydrodynamics to earlier times and more anisotropic systems~\cite{Florkowski:2017olj}. One proposal to potentially extend the validity of hydrodynamics goes by the name of \textit{hydrodynamic attractors} \cite{Heller:2015dha}, which represent emergent constitutive relations away from equilibrium. Over the course of the last few years, the emergence of such hydrodynamics attractors has been firmly established in a variety of effectively 0+1 dimensional microscopic models which feature a rather high degree of symmetry 
\cite{Romatschke:2017vte,Strickland:2017kux,Strickland:2018ayk,Jaiswal:2019cju,Blaizot:2019scw,Almaalol:2020rnu,Heller:2020anv,Du:2020zqg}, and different theoretical~\cite{Romatschke:2017acs,Behtash:2017wqg,Denicol:2017lxn,Denicol:2019lio,Kurkela:2019set,Blaizot:2020gql,Behtash:2020vqk,Du:2021fok,Chattopadhyay:2021ive} and phenomenological aspects of the emergence of hydrodynamic attractors~\cite{Kurkela:2018wud,Giacalone:2019ldn,Jankowski:2020itt,Coquet:2021lca} have been explored. See~\cite{Soloviev:2021lhs} for a recent review.

Despite efforts in this direction~\cite{Romatschke:2017acs,Ambrus:2021sjg}, the generalization of the concept of hydrodynamic attractors to higher dimensional systems represents an outstanding challenge. With this as a motivation, it was recently pointed out that the hydrodynamic attractor can be viewed as a dimensionality reduction in a space of natural observables associated with nuclear collision problems~\cite{Heller:2020anv}. From this perspective, the number of relevant degrees of freedom required to describe the evolution of a system is effectively lowered due to a rapid memory loss of initial conditions. Viewing thermalization in this way enables to approach the study of thermalization in a data-driven way and borrow methods from data science. 

In the present work we adopt the approach of~\cite{Heller:2020anv} to study the evolution of a 0+1D Bjorken flow in an effective kinetic theory (EKT) of pure glue QCD~\cite{Kurkela:2018wud,Giacalone:2019ldn} with highly anisotropic color glass condensate~\cite{Gelis:2010nm,Iancu:2003xm} initial conditions. Building on the methodology developed in \cite{Heller:2020anv}, we use principal component analysis~(PCA) to study the dimensionality reduction for a set of observables that describe the thermalization process at weak coupling. The mechanism behind the information loss can come from the expansion at early time or interactions at late times~\cite{Blaizot:2017ucy,Kurkela:2019set}. We also look more closely at the late time behaviour to gain information on transient non-hydrodynamic contributions to the pressure anisotropy, and compare our results in pure glue QCD kinetic theory to previous studies in conformal Relaxation Time Approximation (RTA)~\cite{hellerHydrodynamizationKineticTheory2018, hellerHowDoesRelativistic2018}. 

The paper is organized in the following way: In Sec.~\ref{sec-setup}, we introduce our model of the initial state, along with the effective kinetic theory of pure glue QCD describing the non-equilibrium dynamics of the QGP. Subsequently, in Sec.~\ref{sec-earlytime}, we perform a PCA of the evolution for different initial conditions to study information loss and dimensionality reduction over the course of the thermalization process of the pre-equilibrium QGP. Sec.~\ref{sec-eq} is devoted to a more detailed study of the evolution towards equilibrium, focusing on the exponential decay of  variance of the pressure anisotropy in the QGP. We conclude this paper in Sec.~\ref{sec-outlook} with a short summary of our most important findings and an outlook on potential future works.

\section{Setup \label{sec-setup}}

We use the Color Glass Condensate~(CGC) effective theory of high-energy QCD~\cite{Gelis:2010nm,Iancu:2003xm}, to evaluate the phase-space distributions of gluons produced in the collision. Due to physical differences and uncertainties in modeling the initial state, this gives rise to a multi-dimensional parameter space of early time phase-space distributions of gluons. They serve as initial conditions for EKT, which we subsequently utilize to evolve expanding nuclear matter into the hydrodynamic phase.

\subsection{Color Glass Condensate initial conditions}
\label{sec-CGC}
We will compute the initial spectrum of gluons based on the $k_T$-factorization formula ~\cite{Blaizot:2010kh, Lappi:2017skr} 
\begin{eqnarray}
\label{eq:spectrum}
&&\frac{dN_{g}}{d^2\bt d^2\Pt dy}= \frac{g^2 N_c}{4\pi^5 \Pt^2 (N_c^2-1)} \\ 
&&\int \frac{d^2\kt}{(2\pi)^2}~\Phi_{A}(\bt+\frac{\bt_{0}}{2},\kt)\Phi_{B}(\bt-\frac{\bt_{0}}{2},\Pt-\kt)\;. \nonumber
\end{eqnarray}
where $\frac{dN_{g}}{d^2\bt d^2\Pt dy}$ describes the transverse momentum $(\Pt)$ spectrum of gluons produced per unit rapidity $(y)$ and transverse area $(\bt)$. By $N_c=3$ we denote the number of colors, $g$ is the Yang-Mills coupling, $\bt_{0}$ denotes the impact parameter of the nucleus-nucleus collision and $\phi_{A/B}(\bt,\kt)$ is the un-integrated gluon distribution from each nucleus (A or B). We employ a particularly simple parametrization of the nuclear gluon distributions, due to Golec-Biernat and Wusthoff (GBW)~\cite{Golec-Biernat:1998zce}, for which
\begin{eqnarray}
\label{eq:UGD_GBW}
\Phi_{A, B}(\bt,\kt) = 4\pi^2 \frac{(N_c^2-1)}{g^2N_c} \frac{\kt^2}{Q_{\rm A, B}^{2}(\bt)} \exp \left( \frac{-\kt^2}{Q_{A, B}^{2}(\bt)} \right)
\end{eqnarray}
where $Q_{A,B}^{2}=Q_{s}^2(\bt)$ denotes the (adjoint) saturation scale~\cite{Albacete:2014fwa} for nucleus.

Based on Eqns.~(\ref{eq:spectrum}) and (\ref{eq:UGD_GBW}), the initial gluon spectrum in the GBW model can then be expressed analytically as 
\begin{eqnarray}
\label{eq-spectrumT}
&&\frac{dN_{g}}{d^2\bt d^2\Pt dy}
=\frac{(N_c^2-1)}{g^2 N_{c} \pi^2 \mathbf{P}^2 }\\ 
\nonumber
&&\frac{x^2 e^{-\Pt^2/Q^2}}{Q^2}  \left(\frac{\Pt^4x^2+\Pt^2(1-x^2)^2Q^2+2x^2Q^4}{(1+x^2)^4}\right)
\end{eqnarray}
where $Q \equiv \sqrt{Q_A^2 + Q_B^2}$ and $x \equiv \max\left(\frac{Q_B}{Q_A},\frac{Q_A}{Q_B}\right)$
denote the root mean square average and the ratio of the saturation scales. Since the saturation scales $Q_{A/B}^{2}$ are generically proportional to local density of nuclear matter, we can anticipate that for different collision geometries the ratio $x=Q_{A}/Q_{B}$ will vary, and we will consider variations $1 \leq x \leq 6$ in the following.

Since the transverse spectrum in Eq.~(\ref{eq-spectrumT}) features a $\propto 1/\Pt^2$ infrared behavior, soft observables, such as screening mass~$m_{D}$ and scattering rate~$g^2T^{*}$ in Eq.~(\ref{eq-deybemass}),
are infrared divergent. However, this divergence can be regulated by non-linear effects, going beyond the factorization formula, see e.g. the discussion in ~\cite{Blaizot:2010kh}. Within our study, we will simply model this by introducing an additional infrared regulator, by virtue of the replacement 
$\frac{1}{\Pt^2} \to \frac{\Pt^2}{(\Pt^2+m^2)^2}$
in the first factor of Eq.~(\ref{eq-spectrumT}), which ensures that $m_{D}$ and $g^2T^{*}$ in Eq.~(\ref{eq-deybemass}) remain finite, while the initial energy density is not affected by the regulator as long as $\text{max}(Q_A^2,Q_B^2) \gg m^2$. Since corrections to the factorization formula become important for $\Pt \lesssim \min(Q_A,Q_B)$, in the following we choose the infrared regulator as 
$\mu \equiv \frac{m}{Q}= \frac{\min(Q_A,Q_B)}{2Q}$. We further note that due to the particular simplicity of the GBW model, the initial spectrum in Eq.~(\ref{eq-spectrumT}) exhibits and exponential decay at high momentum $\Pt \gtrsim Q$, whereas a more realistic parametrization should give rise to a $Q^4/\Pt^4$ power-law tail~\cite{Dumitru:2001ux}. However, since the energy density of the system is dominated by momenta $\Pt \sim Q$, we believe that the model is adequate to study the thermalization of the bulk QGP, while neglecting the impact of high-energy degrees of freedom.

Based on the transverse momentum spectrum, the initial phase-space distribution $f(x,p)$ can then be obtained as~\cite{Greif:2017bnr}
\begin{eqnarray}
f_{g}(\tau_0,\bt,\Pt,y-\eta)= \frac{(2\pi)^3}{2(N_c^2-1)} \frac{\delta(y-\eta)}{|\Pt|\tau_0} \frac{dN_{g}}{d^2\bt d^2\Pt dy}
\end{eqnarray}
where $\eta$ denotes the space-time rapidity and we employ these as initial conditions for the subsequent kinetic description at an initial proper time $\tau_0 = 1/Q$, where a quasi-particle description first becomes applicable~\cite{schlichting2019first,Berges:2020thermalization}. While in the high-energy boost-invariant limit, the initial distribution is proportional to $\delta(y-\eta)$, any interactions will immediately broaden the longitudinal momentum distribution and we therefore consider a smearing form of the delta function 
\begin{eqnarray}
\delta(y-\eta) \to \frac{1}{\sqrt{2\pi}\sigma}\exp\left(-\frac{(y-\eta)^2}{2\sigma^2}\right).
\end{eqnarray}
in the initial conditions. We treat the width of longitudinal rapidity distribution $\sigma$ as the second free parameter in the initial conditions and consider variations in the range $0.05<\sigma<0.50$ in the following.

Starting from the phase-space distribution, the energy-momentum tensor $T^{\mu\nu}$ is defined as a its second moment
\begin{eqnarray}
T^{\mu\nu}=\nu_{g} \int \frac{d^3p}{(2\pi)^3} \frac{p^{\mu}p^{\nu}}{p} f_{g}(\tau,\bt,\Pt,y-\eta)\;.
\end{eqnarray}where $\nu_{g}=2(N_c^2-1)$ denotes the degeneracy factor for gluons. Specifically, one may evaluate the energy density $\epsilon$, the transverse  and the longitudinal pressure $P_{T/L}$ in Milne coordinates as
\begin{subequations}
\label{eq.Tmunufromf}
\begin{eqnarray}
\epsilon&=&T^{\tau\tau}~~
=\int \frac{d^2\Pt}{(2\pi)^2} \frac{dp_{\|}}{(2\pi)}~p~\nu_{g} f_{g}(\tau,\bt,\Pt,p_{\|}),\\
P_T&=&T^{ii}/2
=\int \frac{d^2\Pt}{(2\pi)^2} \frac{dp_{\|}}{(2\pi)} \frac{|\Pt|^2}{2p} \nu_{g} f_{g}(\tau,\bt,\Pt,p_{\|}),\quad\\
P_L&=&\tau^{2}T^{\zeta\zeta}
=\int \frac{d^2\Pt}{(2\pi)^2} \frac{dp_{\|}}{(2\pi)} \frac{p_{\|}^{2}}{p} \nu_{g} f_{g}(\tau,\bt,\Pt,p_{\|}),\quad
\end{eqnarray}
\end{subequations}
where we have re-expressed the phase-space distribution $f_g$ in terms of the momentum variables $p=p^{\tau}=\sqrt{\Pt^2+p_{\|}^2}$ and $p_{\|} = \tau p^{\eta}=|\Pt| \mathrm{sinh}(y-\eta)$. They denote the total and the longitudinal momentum in the co-moving frame.

When considering variations of the parameters $x$ and $\sigma$, we adjust the value of the average saturation scale $Q$, to keep the initial transverse pressure per unit rapidity $g^2 \tau_0 P_T$ (which is independent of the coupling~$g$) constant in physical units. Hence the parameter space of the initial conditions is two dimensional and spanned by the ratio of the nuclear saturation scales~$x$ and the longitudinal smearing width~$\sigma$.\\

\subsection{Effective kinetic theory of pure glue QCD}
\label{sec-YM}
We model the evolution of excited nuclear matter using kinetic description. We follow previous works~\cite{Kurkela:2015qoa,Kurkela:2018vqr,Almaalol:2020rnu} and treat the system as longitudinally boost-invariant and locally homogeneous in the transverse plane. The dynamics reduces to a (0+1)-dimensional problem and is governed by an effective kinetic for pure glue QCD~\cite{Arnold:2002zm}
\begin{eqnarray}
&&\left(\frac{\partial}{\partial \tau} - \frac{p_{\|}}{\tau}\frac{\partial}{\partial p_{\|}} \right) f_g(\tau, \Pt,p_{\|})\\
&&=-C^{{2\leftrightarrow2}}_g[f_{g}](\tau, \Pt,p_{\|})-C^{{1\leftrightarrow2}}_g[f_{g}](\tau, \Pt,p_{\|})\;, \nonumber 
\end{eqnarray}
where $C^{{2\leftrightarrow2}}_g[f]$ denotes the leading order elastic collision integral for gluons  and $C^{{1\leftrightarrow2}}_g[f]$ is the inelastic collision integral that describes emission/absorption of gluon radiation
\begin{widetext}
\begin{subequations}
\begin{eqnarray}
\nonumber
C^{{2\leftrightarrow2}}_g[f](p)
&=&\frac{1}{2 \nu_{g}2 E_p}
\int\frac{d^3p_2}{(2\pi)^3 2E_{p_2}} \frac{d^3p_3}{(2\pi)^3 2E_{p_3}} \frac{d^3p_4}{(2\pi)^3 2E_{p_4}}
(2\pi)^4\delta^{(4)}(p+p_2-p_3-p_4)
|\mathcal{M}_{gg\rightarrow gg}(p,p_2|p_3,p_4)|^2\\
&\times&\left(f(p)f(p_2)(1+f(p_3))f(1+f(p_4))-f(p_3)f(k_4)(1+f(p))f(1+f(p_2))\right),\\
\nonumber
C^{{1\leftrightarrow2}}_g[f](p)
&=&\frac{1}{2}\int_{0}^{1}dz\left\{\frac{d\Gamma_{gg}^{g}}{dz}(p,z)\left[f(p)(1+f(zp))(1+f(\bar{z}p)))-f(zp)f(\bar{z}p)(1+f(p))\right]\right.\\
&-&\left.\frac{1}{z^3}\frac{d\Gamma_{gg}^{g}}{dz}\left(\frac{p}{z},z\right)
\left[f\left(\frac{p}{z}\right)(1+f(p))\left(1+f\left(\frac{\bar{z}}{z}p\right)\right)-f(p)f\left(\frac{\bar{z}}{z}p\right)\left(1+f\left(\frac{p}{z}\right)\right)\right]
\right\}.
\end{eqnarray}
\end{subequations}
\end{widetext}
We note that the matrix element $|\mathcal{M}_{gg\rightarrow gg}(p,p_2|p_3,p_4)|^2$ for elastic scattering is self-consistently screened following the prescription of~\cite{Kurkela:2018oqw}, and that the effective inelastic rates $\frac{d\Gamma_{gg}^{g}}{dz}(p,z)$ account for the Landau-Pomeranchuk-Migdal~effect~\cite{Landau:1953gr,Landau:1953um,Migdal:1955nv} via an effective vertex resummation~\cite{Arnold:2002zm}. Details of the algorithms and numerical implementations can be found in~\cite{Du:2020dvp}. Early initialization requires a fine discretization of the momentum space variables for the EKT simulations, and if not stated otherwise we employ  $N_{p}=1024$ and $N_{\cos(\theta)}=512$ to discretize $p$ and $\cos{\theta}=\frac{p_{\|}}{p}$ in the range between $p_{\rm min}/(g^2\tau_0P_T)^{1/3}=0.01$ and $p_{\rm max}/(g^2\tau_0P_T)^{1/3}=20$. 

Finally, note that in pure glue QCD kinetic theory the interaction strength $g$ and the number of colors $N_{c}$ enter the Boltzmann together as the 't Hooft coupling~$\lambda$~\cite{AbraaoYork:2014hbk}
\begin{equation}
\lambda \equiv g^2 N_{c}.
\end{equation}
Therefore, in the following we will express the interaction strength in terms of $\lambda$.

\section{Early time dynamics, attractors \& dimensionality reduction
\label{sec-earlytime}}

\subsection{Hydrodynamization \& Emergence of attractors}

\begin{figure}
    \centering
     \includegraphics[width=0.5\textwidth]{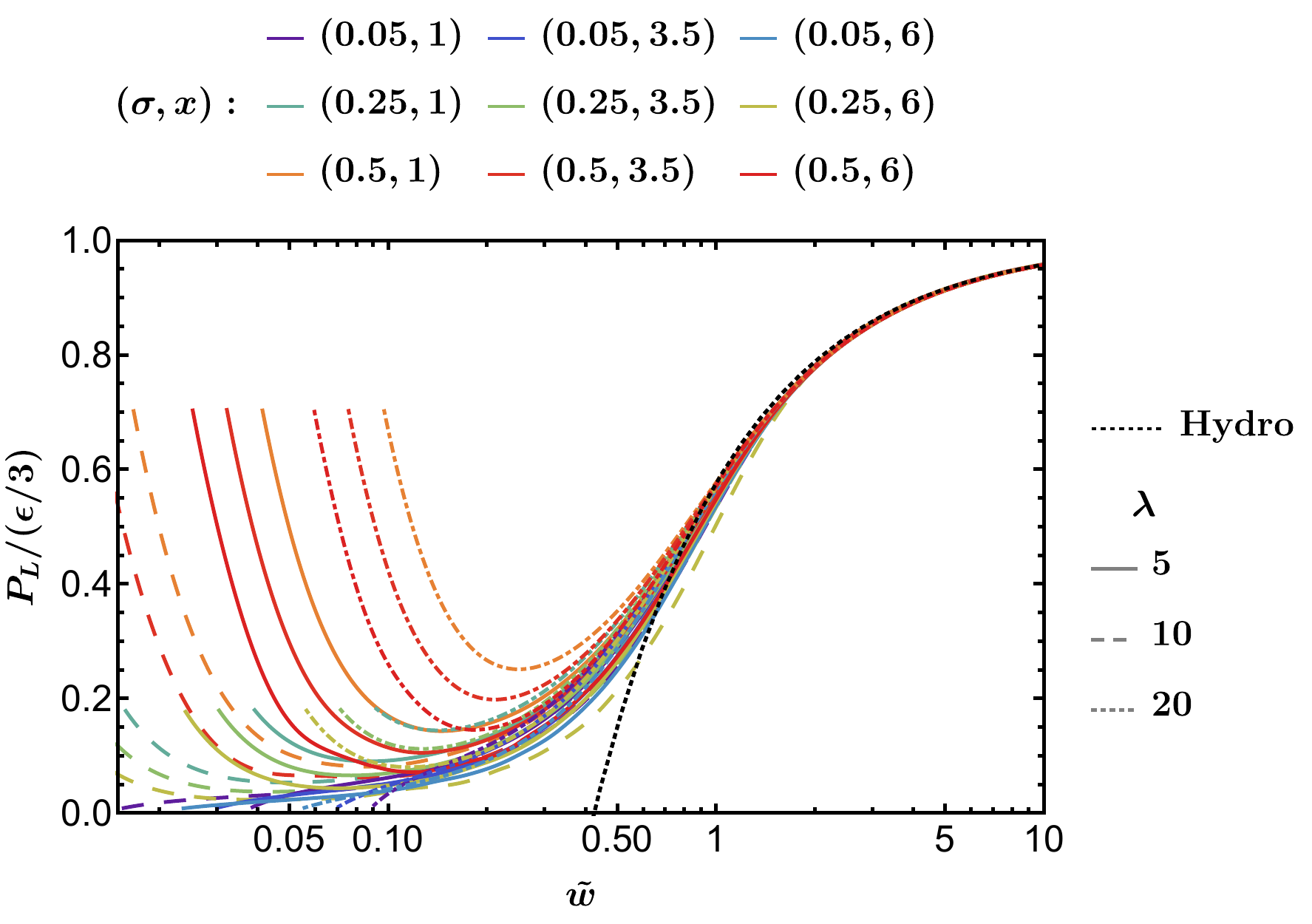}
    \caption{Evolution of the ratio of the longitudinal pressure to what would the equilibrium pressure $P_L/(\epsilon/3)$ as a function of the conformal scaling variable $\tilde{w}$ in Eq.~(\ref{eq:wTilde}). Different curves show results from pure glue QCD kinetic theory simulations for different initial conditions $(\sigma,x)$ at three different coupling strength $\lambda$=5,10,20.}
    \label{fig:ploe} 
\end{figure}

One prominent indicator of transition to hydrodynamics is the longitudinal pressure over what would be the equilibrium pressure ratio $P_L/(\epsilon/3)$ or any of its closely related variants. We present it in Fig.~\ref{fig:ploe} as a function of the universal time scale~\cite{Heller:2011ju,Keegan:2015avk,Heller:2016rtz,Du:2020zqg}
\begin{eqnarray}
\label{eq:wTilde}
\tilde{w}=\frac{\tau (\epsilon+P)}{4\pi\upeta}.
\end{eqnarray}
where $\upeta$ is the shear viscosity and $P=\epsilon/3$ is the thermodynamic pressure for a conformal system. It can be thought of as measuring the physical proper time in units of an effective equilibrium relaxation time 
\begin{equation}
\tau_R=\frac{4\pi\upeta}{\epsilon+P} \equiv \frac{1}{T} \frac{\upeta}{s}/\frac{1}{4\pi},
\end{equation}
where $T$ is an effective temperature of non-equilibrium system determined by~$\epsilon = \nu_{g}\frac{\pi^2 T^4}{30}$ and $s$ is the entropy density. While the overall scaling with $T$ follows on dimensional grounds for a conformal system, the normalization factor of $1/4\pi$ is conventional and inspired by the strong coupling result $\upeta/s=1/4\pi$ of~\cite{Policastro:2001yc,Kovtun:2004de}.

Different colored curves in Fig.~\ref{fig:ploe} show to the results obtained for different ratios of the saturation scales $x$ and different longitudinal smearing width $\sigma$. Different dash styles correspond to the evolution for three different coupling strengths $\lambda=5,10,20$, for which the relevant viscosities $\upeta/s \sim \mathcal{O}(1)$ as summarized in Tab.~\ref{tab:viscosity} in the Appendix.

Starting from the initial conditions, one observes a significant variation of $P_L/(\epsilon/3)$, where small smearing parameters $\sigma$ feature a highly anisotropic initial distribution, with
almost vanishing longitudinal pressure. Since in all cases the initial time $\tau_0=1/Q$ is kept fixed, the curves for different simulation parameters $x,\sigma,\lambda$ start at different values of $\tilde{w}$; in this way larger saturation scale ratios $x$ also tends to provide relevantly more anisotropic initial distribution.
During the initial stages the evolution is dominated by the longitudinal expansion of the system~\cite{Blaizot:2017ucy,Kurkela:2019set}, such that the ratio of $P_L/(\epsilon/3)$ undergoes a quick memory loss and at intermediate times approaches a small value irrespective of the initial conditions. Eventually, on a time scale $\tilde{w}\simeq 1$, all the different curves merge towards the hydrodynamic limit, which to first order in gradients can be universally expressed as~\cite{Heller:2016rtz}
\begin{eqnarray}
\label{eq-visvoushydro}
\left(\frac{P_L}{\epsilon/3}\right)_{\rm hydro}=1-\frac{16}{3}\frac{\upeta}{(\epsilon+P)\tau}
=1-\frac{4}{3\pi\tilde{w}}\;.
\end{eqnarray}

While the emergence of a pressure attractor at $0.1<\tilde{w}<1$ clearly indicates an early reduction of the variance due to the initial free-streaming expansion dynamics, it is equally important to point out that the full convergence to a common attractor only occurs at later times  $\tilde{w}> 1$, when the system relaxes towards hydrodynamics.

\subsection{Principal component analysis}
\label{sec-PCA}

Next, in order to better understand the emergence and properties of the attractor in pure glue QCD kinetic theory, we will scrutinize the evolution further by including additional observables, which go beyond probing the bulk anisotropy of the system. Beyond the anisotropic pressure $P_L$ and $P_T$ in Eqns.~(\ref{eq:pca_observables}), we will study the screening mass $m_D^2$ and the collision rate $g^2T_{*}$
\begin{eqnarray}
\label{eq-deybemass}
m_D^2&=&\frac{4g^2}{d_A}\int \frac{d^3p}{(2\pi)^3}\frac{\nu_g C_Af_g(\vec{p})}{2p}\overset{(eq)}{=}\frac{g^2N_cT_{\rm eq}^2}{3}\\
T_{*}&=&\frac{g^2}{d_A m_D^2}\int \frac{d^3p}{(2\pi)^3}\bigg\{\nu_g C_Af_g(\vec{p})(1+f_{g}(\vec{p}))\bigg\}\overset{(eq)}{=}T_{\rm eq}.\nonumber
\end{eqnarray}
Since $m_{D}^2$ and $g^2T^{*}$ govern the strength of elastic and inelastic interactions in pure glue QCD kinetic theory, they present the most natural quantities to include in a weak coupling analysis.

Since all of of the above quantities $P_L,P_T,m_D^2,T_{*}$ are dimensionful, it is natural to consider dimensionless ratios along with one single quantity that is sensitive to the overall energy scale. Specifically, for our analysis, we choose to consider $P_{L}/(\epsilon/3)$ and normalize $m_D^2,T_{*}$ by the transverse pressure $P_T$. This gives us the following dimensionless quantities
\begin{subequations}\label{eq:pca_observables}
\begin{align}
        \bar{P}_L &\equiv \frac{P_L}{\epsilon/3},\\
    \bar{m}_D^2 &\equiv \frac{m_D^2}{(90 P_T/\pi^2 \nu_g)^{1/2}g^2N_c/3}, \\
    \bar{T}_* &\equiv \frac{T_*}{(90 P_T/\pi^2 \nu_g)^{1/4}}. \end{align}
All of them by construction approach unity in thermal equilibrium. While initially, the transverse pressure $P_T$ is kept constant for different initial conditions, differences in the kinetic evolution gives rise to variations of the overall energy scale, which we monitor in terms of the dimensionless quantity  
\begin{align}
\label{eq:pressureest}
    \bar{P}_T &\equiv \frac{\tau^{4/3} P_T}{1/3 (2P_{T0}\tau_0)^{8/9} (\frac{\pi^2}{30} \nu_{g})^{1/9} (4 \pi \frac{\upeta}{s} )^{4/9}C_{\infty} }.
    \end{align}
\end{subequations}
where we employ $C_{\infty}=0.98$~\cite{Giacalone:2019ldn}. By following the arguments of~\cite{Giacalone:2019ldn,Du:2020zqg}, the denominator in Eq.~(\ref{eq:pressureest}) provides an estimate for the late time asymptotic value of $\tau^{4/3} P_T$, such that for typical initial conditions this quantity can again be expected to be close to unity at late times.

\begin{figure}[t!]
    \centering
    \includegraphics[width=0.5\textwidth]{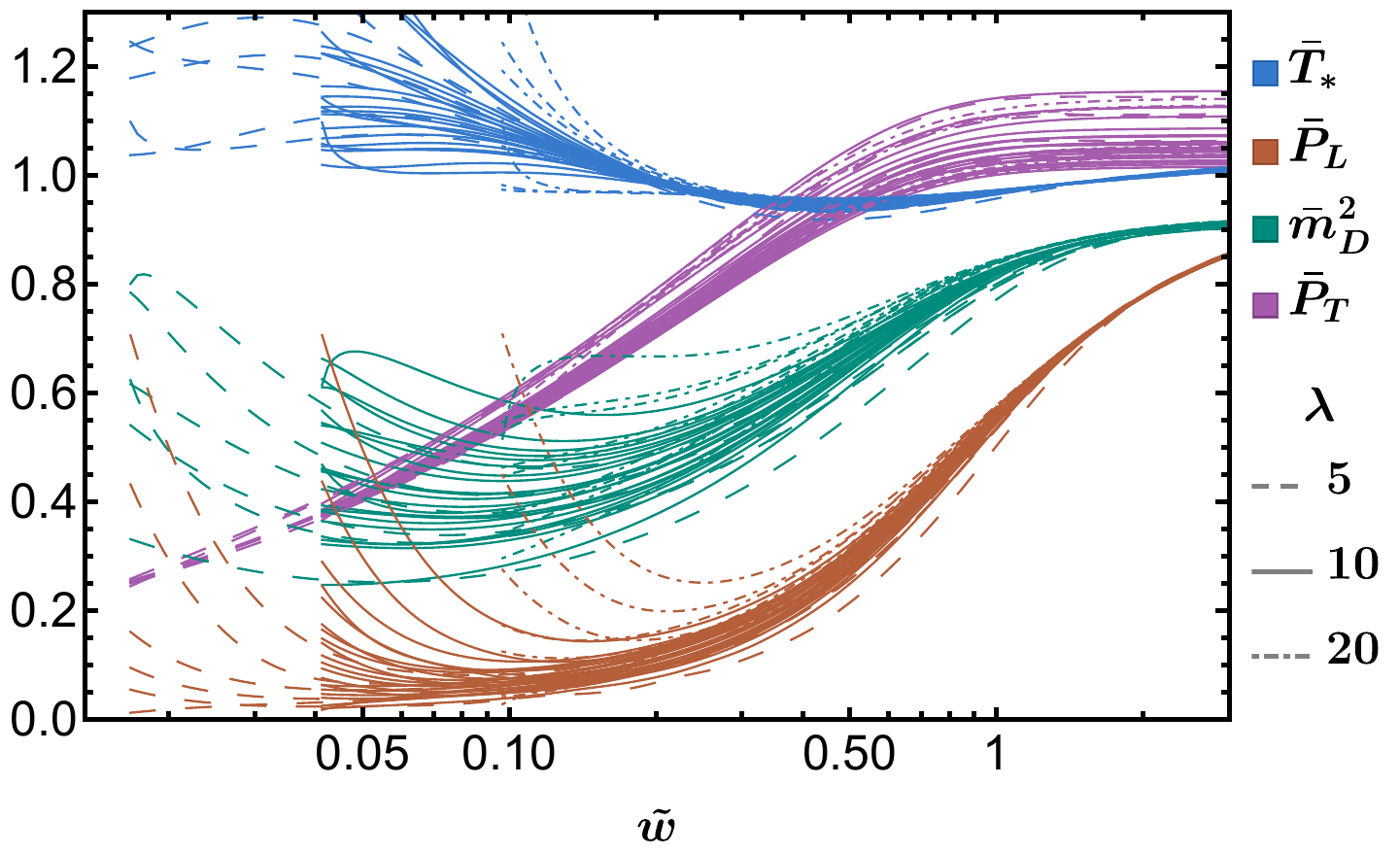}
    \caption{Evolution of the observables $\bar{P}_T,\bar{P}_L,\bar{m}_D^2,\bar{T}_{*}$ as a function of the universal time scale $\tilde{w}$ for a variety of different initial conditions at three different coupling strengths $\lambda=5,10,20$. The observable $\bar{P}_T$ has been chosen to contain the information on the overall energy scale and features the only sizable variation at late times. Conversely, all other observables are insensitive to the overall energy scale  and show an attractor behaviour. \label{fig:cgc_observables}}
\end{figure}

We can directly verify these expectations in Fig.~\ref{fig:cgc_observables}, where we present the evolution of the observables $\bar{P}_T,\bar{P}_L,\bar{m}_D^2,\bar{T}_{*}$ as a function of the universal time scale $\tilde{w}$ for a variety of different initial conditions at three different coupling strengths $\lambda=5,10,20$. While the self-normalized quantities $\bar{P}_L,\bar{m}_D^2,\bar{T}_{*}$ all converge to a common attractor behavior at late times, the overall energy scale $\bar{P}_T$ develops a $\sim 15\%$ variation over the course of the non-equilibirum evolution of the system.

With the set of observables in Eq.~(\ref{eq:pca_observables}), we can further perform a PCA to extract the most significant contributions in the emergence of the attractor phenomenon, namely the principal components of the attractor. For this purpose, rather than studying these quantities separately, we can also consider these quantities as parametrizing a four dimensional space. Each solution is represented as a vector
\begin{equation}
    \mathbf{X}(\tilde{w}) = \begin{pmatrix} \bar{T}_{*}(\tilde{w}) & \bar{P}_{L}(\tilde{w}) & \bar{m}_{D}^2(\tilde{w}) & \bar{P}_{T}(\tilde{w}) \end{pmatrix}
 \end{equation}
evolving in time. Evidently, to perform a meaningful comparison of the different directions in $\mathbf{X}(\tilde{w})$ space, the units and overall scales of the different components of $\mathbf{X}(\tilde{w})$ must be comparable, which is precisely the reason that the observables have been normalized as in Eq.~(\ref{eq:pca_observables}).

From this point of view, thermalization is achieved through dimensionality reduction. With sufficient variation of the initial phase-space distribution function, the set of initial conditions $\{\mathbf{X}_{0}\}$ collectively spans some volume in the four dimensional space. However, since the initial conditions used here only depend on two parameters, we only expect to span (at most) a two dimensional subspace. Over the course of the thermalization process, the dimensionless variables $\bar{P}_L,\bar{m}_D^2,\bar{T}_{*}$ all approach unity, and only $\bar{P}_T$ which is sensitive to the overall energy scale should have a significant variation at late times, implying that ultimately the solutions will span a one dimensional subspace. 

%%%%%%%

PCA is a simple method to study this process of dimensionality reduction. Given a set of points $\{ \mathbf{X} \}$, the PCA calculates the eigenvectors and eigenvalues of the covariance matrix
\begin{equation}
    \mathbf{C}_{mn} = \operatorname{Cov}(\mathbf{X}_m, \mathbf{X}_n).
\end{equation}
to produce a set of orthonormal vectors such that the first vector points along the direction of the largest variance of the data set, and the subleading vectors are similarly optimized in the space orthogonal to the leading vector. Each vector is associated with an explained variance, quantified by the variance of the set of points in its direction. The number of non-negligible explained variances gives a measure of the dimensionality of the region occupied by the states of interest at a given value of time variable.

\begin{figure*}[th!]
    \centering
    \includegraphics[width=.9\linewidth]{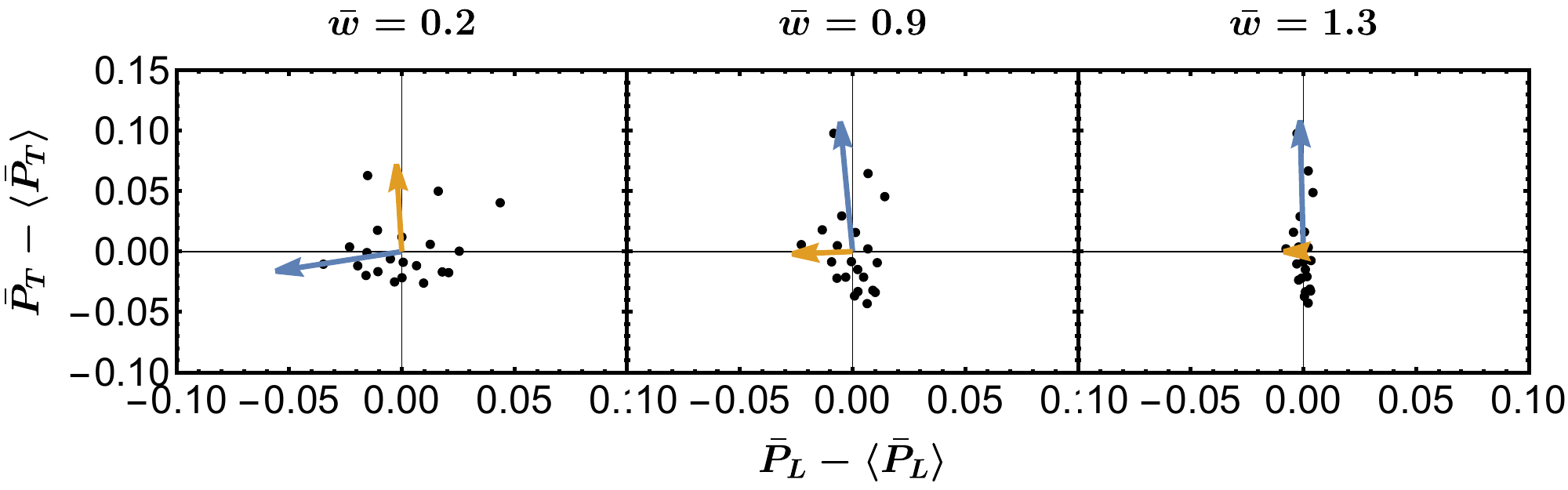}
    \caption{Snapshots of the evolution of an ensemble of different initial conditions in the $\bar{P}_L$,$\bar{P}_{T}$ plane at three different times $\tilde{w}=0.2,0.9,1.3$ for $\lambda=10$. Each point corresponds to the evolution for a particular initial condition; blue and orange arrows indicate the first and second principle components. While initially the points are scattered in the two-dimensional plane, they collapse onto a one dimensional subspace at late times. The length of the vectors is set to three times the square root the of the explained variance for better visibility.}
    \label{fig:pca_illustration}
\end{figure*}

We illustrate this behavior in Fig.~\ref{fig:pca_illustration}, where we show
the distribution of values in the $\bar{P}_L$,$\bar{P}_T$ plane at three different times $\tilde{w}=0.2,0.9,1.3$ of the evolution. Each point in  Fig.~\ref{fig:pca_illustration} corresponds to the evolution for a particular initial condition, while the blue and orange arrows indicate the first and second principle components. While at early times $(\tilde{w}=0.2)$ the different initial conditions cover a two-dimensional subspace, with the largest variations in the $\bar{P}_L$ direction, the effective reduction of the dimensionality of the distribution is clearly visible, as at late times $\tilde{w}=1.3$ all points converge towards a one dimensional manifold oriented along the $\bar{P}_T$ direction.

\begin{figure}
    \centering
    \includegraphics[width=0.44\textwidth]{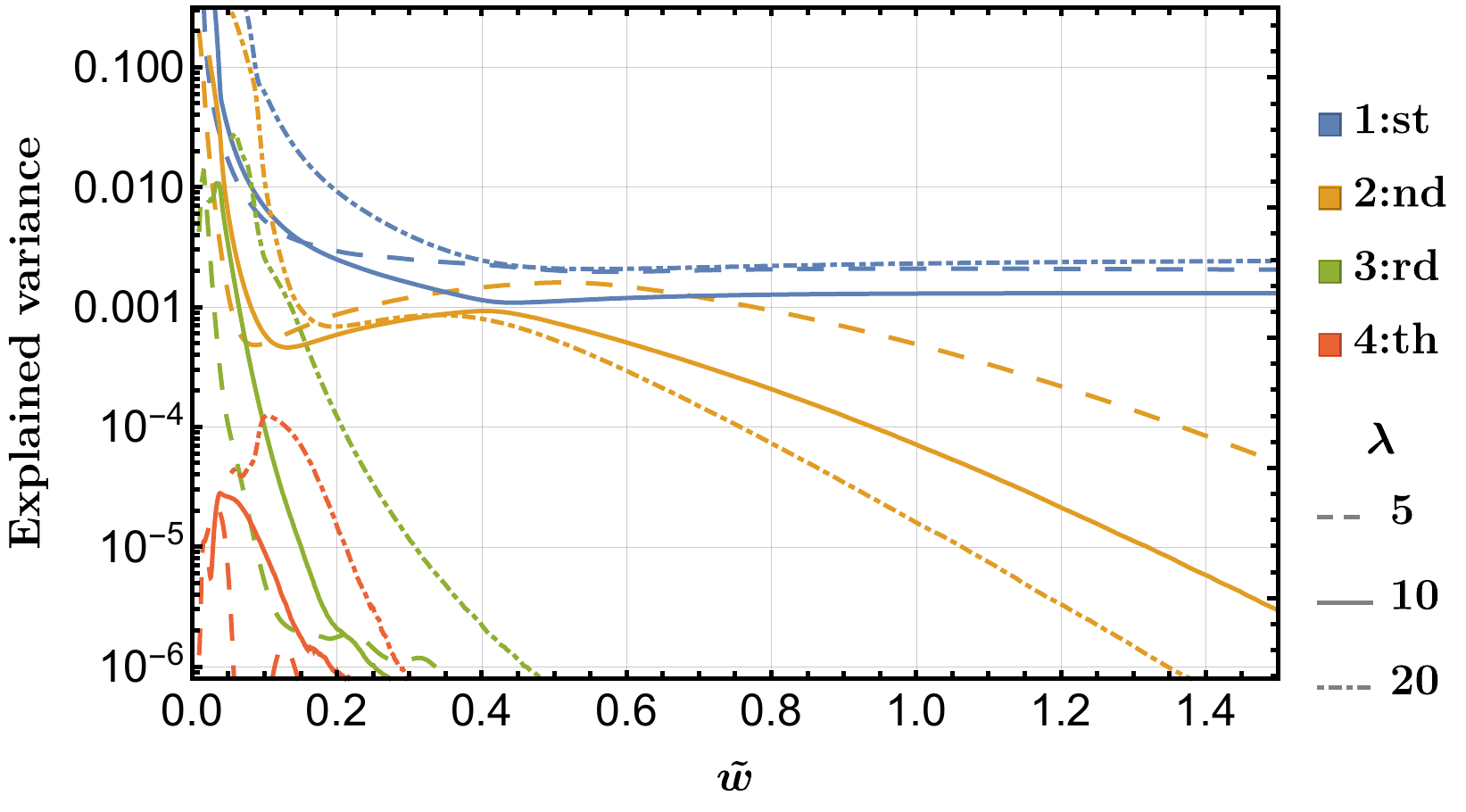}
    \caption{Explained variances for the observables in Fig.\,\ref{fig:cgc_observables}. The colors distinguish between different principal components, ordered by the explained variance. Starting from a complicated behavior at early times, the only relevant contributions at late times are associated with a single dominant component that remains approximately constant and a second component which is approximately exponentially decaying. Separate contributions of the different observables to the these two principal components are shown in Fig.\,\ref{fig:pca_vectors}.
    \label{fig:pca_variance}
    }
\end{figure}

\begin{figure}
    \centering
    \includegraphics[width=0.44\textwidth]{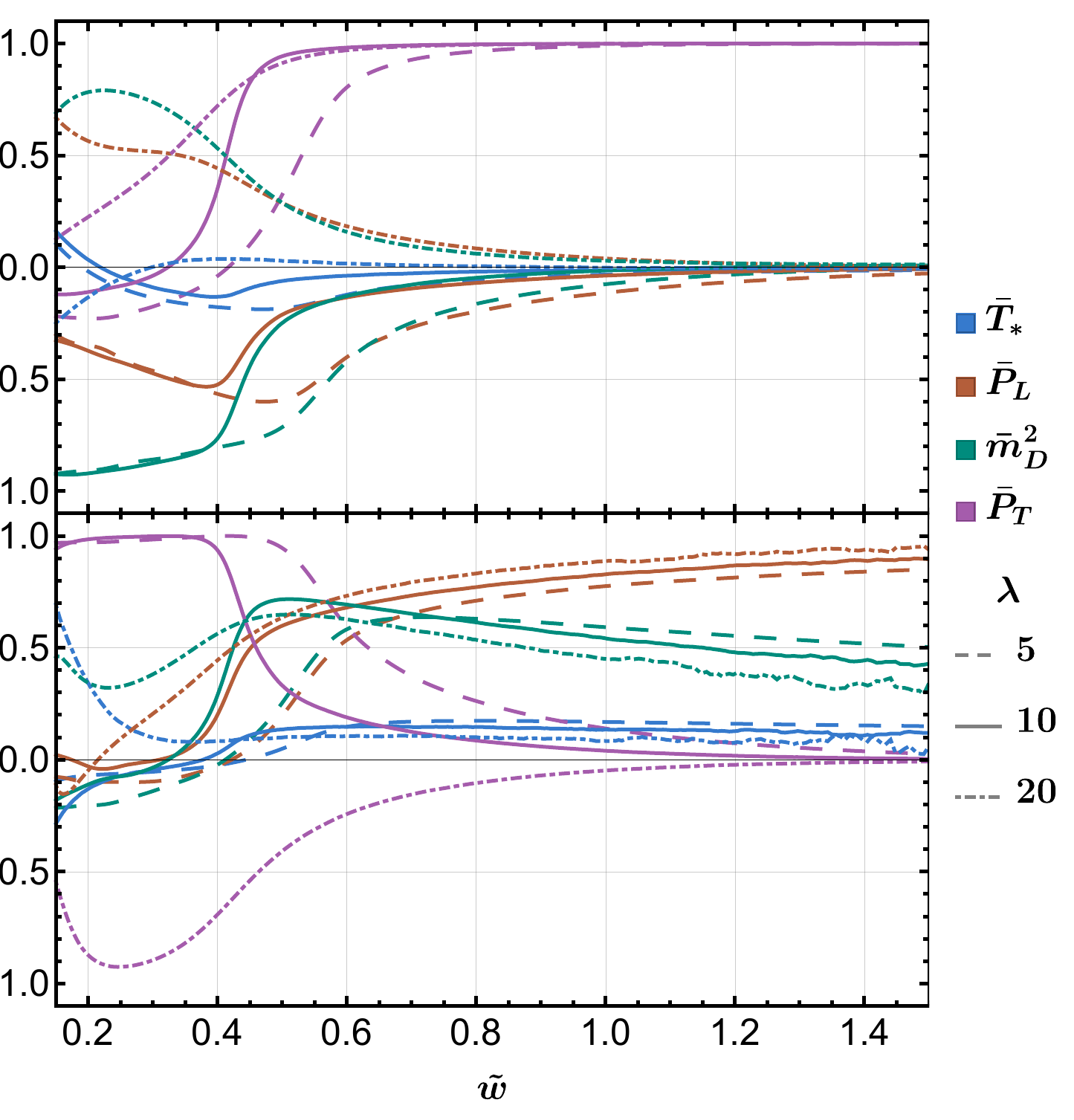}
    \caption{ Decomposition of the leading (top) and first subleading (bottom) principal component vectors. By design, the leading vector is dominated by $\bar{P}_T$, which carries information about the overall energy scale. The second principal component corresponds to an exponentially decaying transient, and characterizes the variation in the other observables. Since no other principal component is sizable, we learn that the observables $\bar{m}_D^2,\bar{T}_{*},\bar{P}_L$ are highly correlated and governed by the same attractor.
    \label{fig:pca_vectors}
    }
\end{figure}

By applying PCA to the set of solutions at each point in the universal time $\tilde{w}$, we can study the dimensionality of the dataset through the explained variances, and identify which directions in the space of observables that show the greatest variance. Our results are compactly summarized in Figs.\,\ref{fig:pca_variance} and \ref{fig:pca_vectors}, where we show the evolution of the explained variances along with the evolution of the composition of two dominant principle component vectors. One clearly observes from Fig.\,\ref{fig:pca_variance}, that at late times, there is indeed only a single dominant component, which is associated with an approximately constant explained variance. Beyond the leading principal component, the explained variance of the second most relevant component decays approximately exponentially at times $\tilde{w}\gtrsim 0.5$.

The decomposition of the first and second principal components are shown in Fig.\,\ref{fig:pca_vectors}. The abrupt shift in behaviour at $\tilde w \approx 0.5$ is correlated with the crossing of explained variances in Fig.\,\ref{fig:pca_variance}, and should be thought of as the two vectors switching identity. At late times the first component is dominated by $\bar{P}_T$, which was chosen to be sensitive to the energy scale of the state. The second component is dominated by $\bar{P}_L$ and $\bar{m}_D^2$.

While Fig.\,\ref{fig:cgc_observables} already shows that the observables $\bar{m}_D^2,\bar{T}_{*}$ also features a similar attractor behaviour as the well-known $\bar{P}_L$, it does not tell us whether these attractors are actually the same. In contrast, the principal component analysis in Figs.\,\ref{fig:pca_variance} and \ref{fig:pca_vectors} reveals that the evolution of the different observables is highly correlated and can be captured with a single principal component.

\section{Evolution towards equilibrium}
\label{sec-eq}
So far we have statistically analyzed the emergence of hydrodynamics attractors starting from CGC initial conditions at very early times $\tilde{w} \ll 1$. While in this case the dynamics is initially dominated by the longitudinal expansion, we performed additional simulations that focus on the late time approach to hydrodynamic behavior to further analyze the memory loss and convergence towards hydrodynamic behavior in pure glue QCD kinetic theory. By initializing the system according to a Romatschke-Strickland type distribution~\cite{Romatschke:2003ms},
\begin{eqnarray}
\label{eq:f0latestart}
f_{g}(\tau_0,\Pt,p_{\|})=\frac{c(\xi_0)}{e^{\frac{\sqrt{\Pt^2+\xi_{0}^{2}p_{\|}^2}}{T_0}}-1}
\end{eqnarray}
for different initial anisotropies $\xi_0=1.25,2.5,5,10$ at initial time $\tilde{w}_0=0.1,0.3$ and $\xi_0=1.1,1.2,1.3$ at initial time $\tilde{w}_0=1,3,10$, we are then able to analyze the effective memory loss at late times for a large range of couplings $\lambda=0.1 - 20$. We note that in all cases, the normalization factor
\begin{eqnarray}
c(\xi_{0})=\frac{2}{1/\xi_{0}^2+\text{arctan}(\sqrt{\xi_{0}^2-1})/\sqrt{\xi_{0}^2-1}}
\end{eqnarray}
is chosen such that the initial energy density $\epsilon(\tilde{w}_0)=\frac{\pi^2}{30}\nu_{g} T_0^4$ remains the same irrespective of the initial anisotropy $\xi_0$; while in order to initialize the simulations at the same $\tilde{w}_0$, the initial proper time is adjusted as $\tau_0=\frac{4\pi \upeta/s}{T_0} \tilde{w}_0$ in Eq.~\eqref{eq:f0latestart} for the respective coupling strength. For completeness, our numerical implementation employs the following discretization: $N_{p}=64$ and $N_{\cos(\theta)}=64$ with $p_{\rm min}/T_0=0.01$ and $p_{\rm max}/T_0=8$.

\subsection{Exponential approach to hydrodynamics at late times}
\label{sec-exp}

In order to investigate direct approach to viscous hydrodynamics at late times, we will focus on the evolution of $P_L/(\epsilon/3)$ in $\tilde{w}$. As we discussed in Sec.~\ref{sec-PCA}, this quantity has a significant contribution to the leading principal component characterising deviations from the attractor, and exhibits a universal late time hydrodynamic behavior determined by Eq.~(\ref{eq-visvoushydro}). 

\begin{figure}[t]
    \centering
    \includegraphics[width=0.44\textwidth]{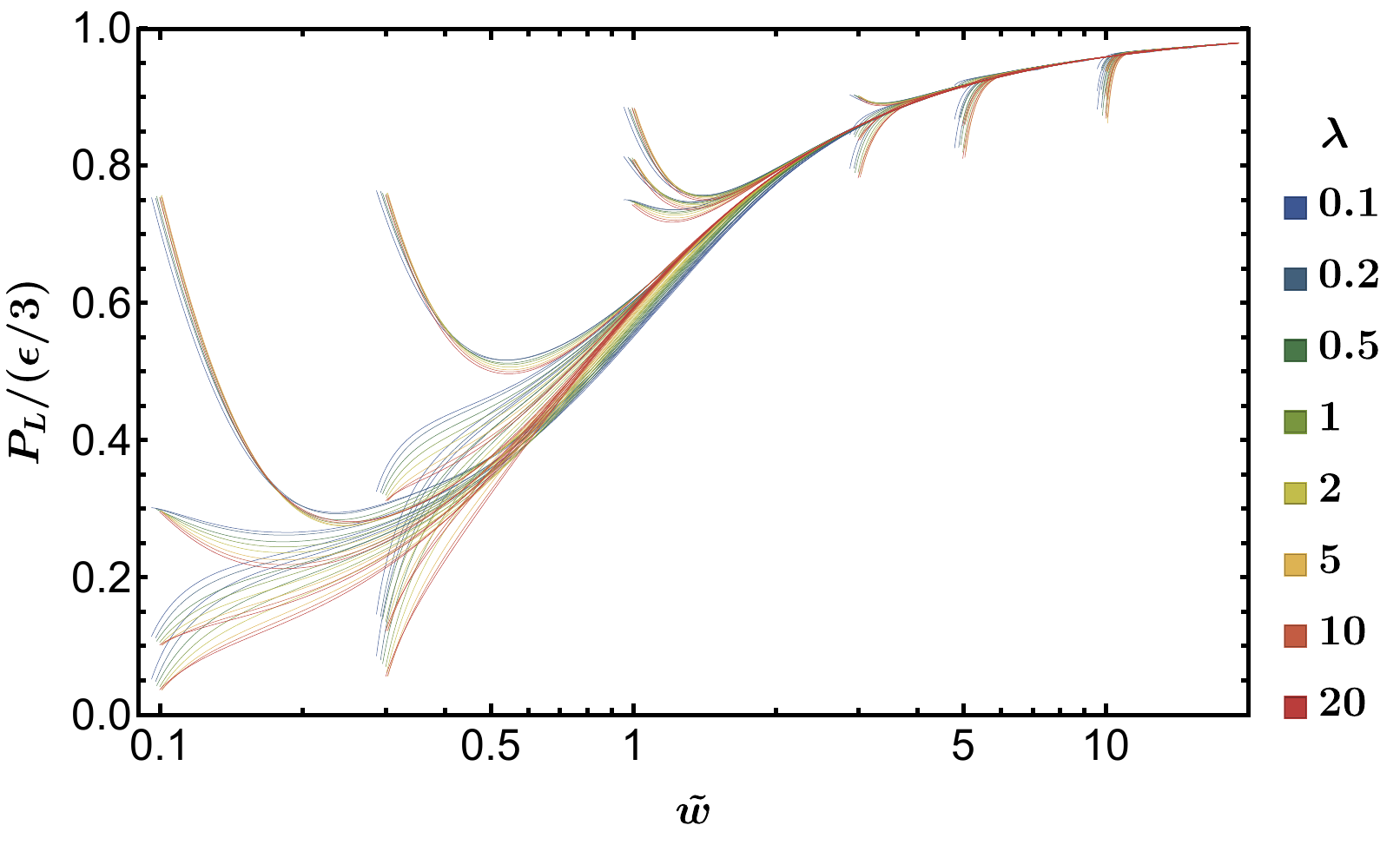}
     \includegraphics[width=0.44\textwidth]{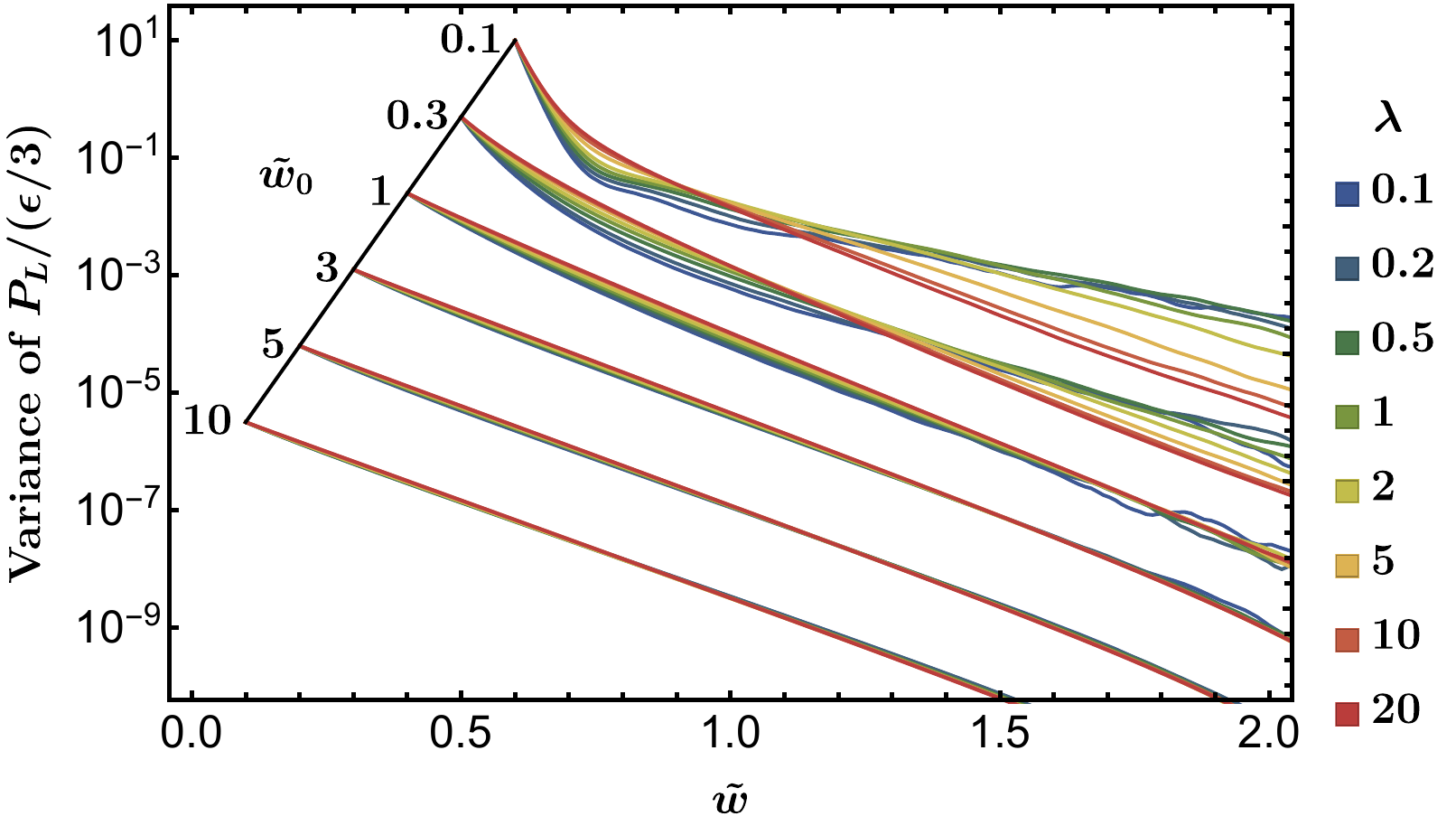}
     \includegraphics[width=0.44\textwidth]{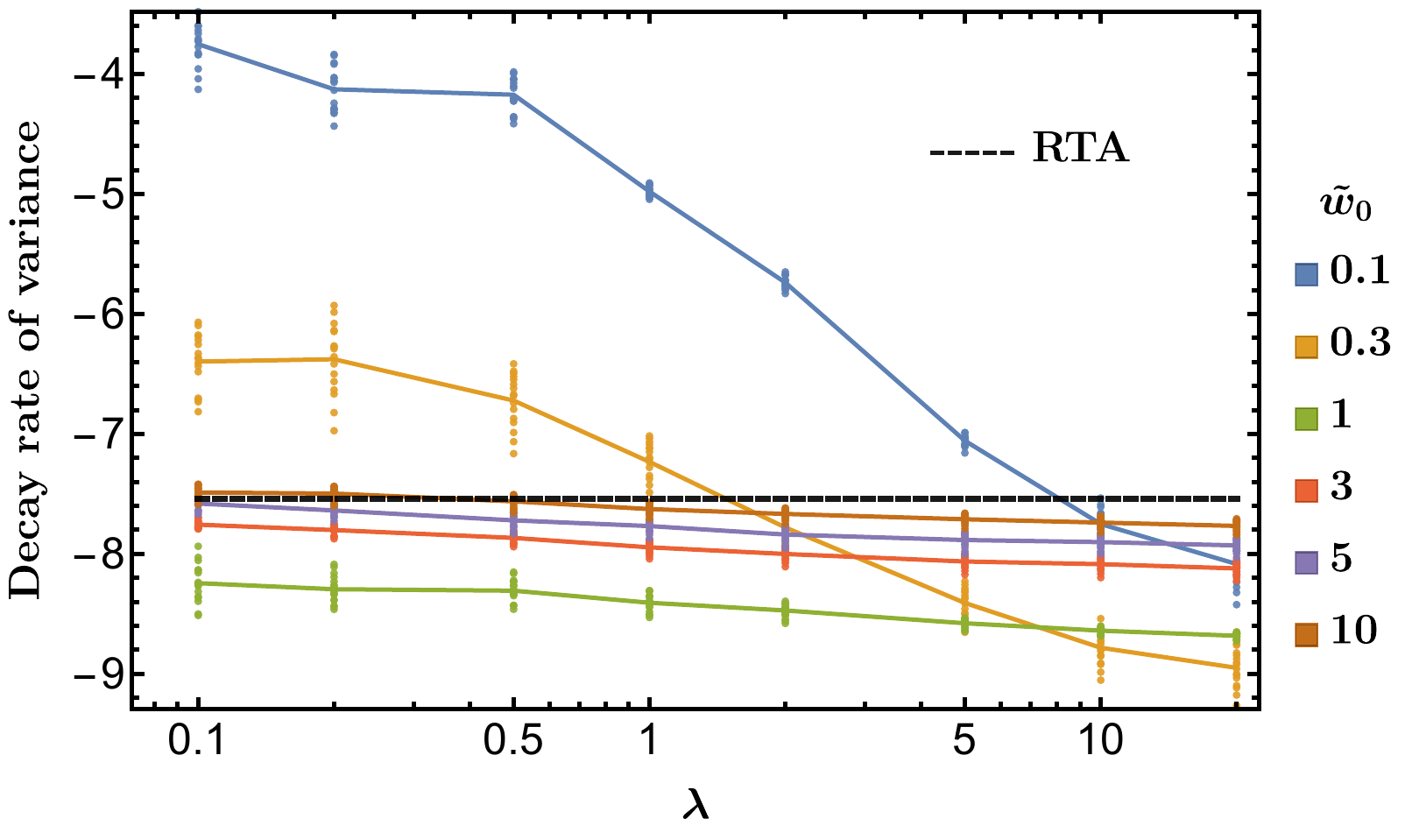}
    \caption{ (top) Evolution of the pressure to what would be the equilibrium pressure ratio $p_L/(\epsilon/3)$. (middle)  Evolution of variance of $P_L/(\epsilon/3)$. The exponential decay is associated to decay of non-hydrodynamic contributions to $P_L/(\epsilon/3)$. (bottom) Exponential decay rates of the variance extracted from the middle figure. Different points are obtained by varying the extraction range and the curve is formed from the mean result.  \label{fig:pLOverE}
    }
\end{figure}

The evolution of the $P_L/(\epsilon/3)$ for the initial conditions~\eqref{eq:f0latestart} is depicted in Fig.~\ref{fig:pLOverE}. Different colored curves correspond to the results for different coupling strength $\lambda=0.1 - 20$, while curves of the same color correspond to different initial conditions $\xi_0$ at different initialization times $\tilde{w}_0$. Irrespective of the coupling strength, one observes that very quickly after the initialization, different initial conditions appear to converge towards a common attractor curve. By careful inspection, one notes that for varying coupling strength slight differences in the attractors persist at intermediate times $\tilde{w}\lesssim3$, before eventually all curves converge towards the same (by construction) hydrodynamic late time behavior~\eqref{eq-visvoushydro}. Note that contributions from the terms second and higher order in derivatives are expected to exhibit residual coupling dependence, which might at least partially explain the slight differences between attractors at earlier times.

In order to further characterize the approach towards an attractor, we compute the variances of $P_L/(\epsilon/3)$ for the different initial conditions $\xi_0$ at each value of the initialization time $\tilde{w}_0$ and coupling strength~$\lambda$. Before we discuss our results in pure glue QCD kinetic theory, shown in the middle panel of Fig.~\ref{fig:pLOverE}, it proves insightful to recall the behavior previously observed in different microscopic models of early-time dynamics of the QGP.

Starting with \cite{Heller:2015dha} (see also~\cite{Basar:2015ava,Aniceto:2015mto} for a more complete discussion), it was understood that in a class of models employing the M{\"u}ller-Israel-Stewart (MIS) approach to embed hydrodynamics in a framework compatible with relativistic causality~\cite{Muller:1967zza,Israel1976Sep,Israel:1979wp}, the late time behavior of $P_L/(\epsilon/3)$ can be elevated into a transseries~\cite{Aniceto:2018bis} of the form 
\begin{equation}
\label{eq.transseries}
    \frac{P_L}{\epsilon/3} = \sum_{n=0}^\infty \frac{b_n}{\tilde{w}^n} + \sum_j e^{-\Omega_{j} \tilde{w}}\tilde{w}^\beta_j\sum_{n=0}^\infty \frac{b_{j,n}}{\tilde{w}^n}.
\end{equation}
Without dwelling into details of this formal expression, the key aspect for us are exponentially suppressed in times effects with decay rates associated with $\Omega_{j}$. In MIS different $\Omega_{j}$ are just integer multiples (due to nonlinear effects) of a single relaxation scale present in this class of theories. The same structure was verified to appear in RTA kinetic theory~\cite{hellerHydrodynamizationKineticTheory2018,hellerHowDoesRelativistic2018,Heller:2021yjh} and in~holography~\cite{Heller:2013fn,casalderreysolanaResurgenceHydrodynamicAttractors2018, anicetoLargePropertimeExpansion2019,Heller:2021yjh}.

Most importantly, the contributions $\Omega_i$ describe the decay of transient non-hydrodynamic contributions to the pressure anisotropy, which in the aforementioned examples can be related to the analytic structure of retarded correlation functions of the energy-momentum tensor in equilibrium~\cite{Kovtun:2005ev,Florkowski:2017olj}. In the conformal RTA kinetic theory the shear viscosity and the relaxation time are related by 
\begin{equation}
\label{eq.RTAtauRetas}
\tau_\text{r}(\tau) = \frac{5\upeta}{sT(\tau)}.
\end{equation}
Following~\cite{Janik:2006gp}, the relaxation time determines the decay rate of non-hydrodynamic contributions in the boost-invariant background as
\begin{align}
\label{eq:RTADecayRate}
    -\int \frac{d\tau}{\tau_\text{r}(\tau)} =  -\int \frac{T(\tau) d\tau}{5 \upeta/s} \approx -\frac{3}{2} \frac{T(\tau) \tau}{5\upeta/s} = -\frac{6}{5}\pi \tilde w,
\end{align}
where we have used the fact that $T(\tau)$ scales as $\tau^{-1/3}$ at sufficiently late times in Bjorken flow and dropped subleading contributions at late time. While Eq.~(\ref{eq:RTADecayRate}) provides the rate of decay of non-hydrodynamic contributions in each individual realization, the variance measures the square of these contributions and thus decays twice as fast.

While the structure of non-hydrodynamic excitations in QCD kinetic theory is generally expected to be rather complicated~\cite{Kurkela:2017xis,Moore:2018mma}, it is nevertheless interesting to investigate to what extent a simple parametrization of the type in Eq.~\eqref{eq.transseries} can describe the convergence towards an attractor. 

To this end, inspection of the evolution of the variances in the middle plot in Fig.~\ref{fig:pLOverE} shows a clear exponential decay of the variance for most values of $\lambda$ and $\tilde w_0$. Clear deviations from an exponential decay are only seen for data initialized at very early times, particularly for small coupling, where subleading contributions to the pressure anisotropy can be expected to be more important.

We extract the effective decay rate $\Omega_{\rm eff}$ of the variance for each choice of $\lambda$ and $\tilde w_0$ by a linear fit in a variety of windows. Extracted values of the decay rate are presented in the bottom plot of Fig.\,\ref{fig:pLOverE}, along with the mean value of those fits and for comparison we also show the decay rate $\frac{12}{5}\pi$ in conformal RTA kinetic theory.

While not perfectly exponential, the data initialized at early times have a smaller effective decay rate at smaller couplings. For the data initialized at late times, which show a clear exponential decay, the decay rate is approximately independent of the coupling strength $\lambda$ (with variations within 5\% as the coupling varies by over two orders of magnitude) and turns out to be rather close to the value in RTA.

While our finding suggests that, just like in RTA, the decay rate in QCD kinetic theory appears to be closely connected to the value of viscosity, we certainly do not have a good explanation of this behavior. We note however, that similar observations were made also at the level of second order transport coefficients~\cite{York:2008rr}. In particular, this implies that the contribution at $\tilde{w}^{-2}$ in Eq.~\eqref{eq-visvoushydro} exhibits very weak residual dependence on the coupling~$\lambda$.

\section{Summary \& Outlook}
\label{sec-outlook}
We have studied the transition to hydrodynamics of Yang-Mills kinetic theory undergoing Bjorken expansion. For initial conditions inspired by the CGC effective theory, we analyzed the subsequent evolution of a natural set of four observables given by Eq.~\eqref{eq:pca_observables} using PCA. Such an analysis may be sensitive to the precise initial conditions used, which is why we studied realistic initial conditions from the CGC effective theory.

Our studies showed that at late times there is a single dominant principal component that describes the variation of the overall energy scale $\bar{P}_T$. Furthermore, the late time evolution of $\bar{P}_L$, $\bar{m}_{D}^2$ and $\bar{T}^{*}$ is highly correlated and can be captured by a second subleading principal component. While at this point it remains a logical possibility that this correlation is an artifact of the considered initial conditions, this could be further explored within a higher dimensional parameter space.

The second principal component represents a transient contribution, which we studied in more detail by considering $\bar{P}_L$ using other initial conditions initialized at different times. A clear exponential decay can be seen at late times by comparing different profiles of $\bar{P}_L$, signaling the presence of an exponential approach to the hydrodynamic attractor at late times. 

Quite surprisingly, for the whole range of the couplings considered, i.e. for $\lambda$ between 0.1 and 20, the decay rate is close to the value predicted by the conformal RTA kinetic theory. What this means in practical terms is that the effective decay rate is simply expressible in terms of the shear viscosity-to-entropy density ratio, see Eq.~\eqref{eq.RTAtauRetas}.

\begin{figure*}[t]
    \centering
     \includegraphics[width=.9\textwidth]{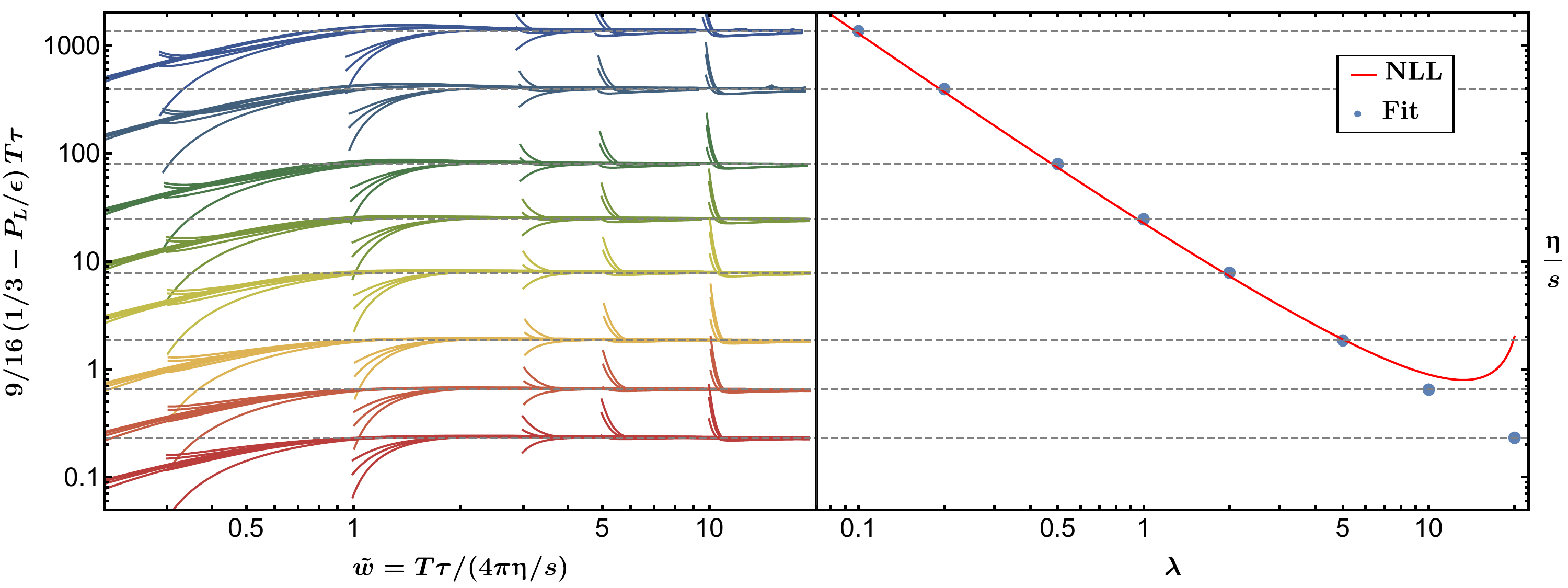}
    \caption{ (left) Extraction of the shear viscosity to entropy density ratio $\upeta/s$ from the late time hydrodynamic behavior of Yang-Mills kinetic theory simulations. (right) Extracted values of $\upeta/s$ as a function of $\lambda=g^2N_c$. See Table\,\ref{tab:viscosity} for numerical values.
    \label{fig:viscosity_extraction}
    }
\end{figure*}

It is an interesting question to ask if the similarities between the EKT and conformal RTA go beyond this crude characteristic. To this end, an in-depth analysis of the RTA kinetic theory in~\cite{hellerHowDoesRelativistic2018} reveals that the exponentially decaying contribution there is not a single excitation, but actually a sum of infinitely many contributions whose occupation numbers are initial condition dependent. While they all are characterized by the same exponential decay rate, their subleading behaviour is distinct yet impossible to disentangle over the short time scales probed in the present project. This indicates that in reality the exponential decay we are reporting here is likely to be understood as an effective description (appropriate for the conformal Bjorken flow with no transverse dynamics) of an underlying more complicated structure.

Beyond further explorations of the surprising similarity in the highly symmetric Bjorken flow, it would also be interesting to compare the evolution of the energy-momentum tensor in a variety of other settings. While first steps in this direction have been reported in~\cite{Kamata:2020mka,Kurkela:2018vqr} for linearized perturbations around Bjorken flow and in ~\cite{Kurkela:2018qeb,Ambrus:2021fej,Ambrus:2021sjg} for systems undergoing both longitudinal and transverse expansion, a particularly clean probe concerns the evolution of energy-momentum (or metric) perturbations in equilibrium. While for conformal RTA kinetic theory the corresponding retarded correlators of the energy-momentum tensor were calculated analytically in~\cite{Romatschke:2015gic}, a dedicated study in QCD kinetic theory would certainly shed further light on the structure and importance of non-hydrodynamic excitations. One can view the results of~\cite{Kurkela:2017xis} obtained in RTA kinetic theory with momentum-dependent relaxation time (see, however,~\cite{Rocha:2021zcw} for a subtlety in this model that emerged after~\cite{Kurkela:2017xis} was completed) as a first step in this direction.

\begin{acknowledgments}

We thank V.~Ambrus, A.~Kurkela, A.~Mazeliauskas, A.~Serantes,  M.~Spali{\'n}ski and B.~Withers for insightful discussions and collaboration on related topics. X.D and S.S acknowledge support by the Deutsche Forschungsgemeinschaft (DFG, German Research Foundation) through the CRC-TR 211 'Strong-interaction matter under extreme conditions'– project number 315477589 – TRR 211. 
V.S acknowledges support from the National Centre for Nuclear Research, Poland. 
The authors also acknowledge computing time provided by the Paderborn
Center for Parallel Computing (PC2) and the National
Energy Research Scientific Computing Center, a DOE
Office of Science User Facility supported by the Office
of Science of the U.S. Department of Energy under Contract No. DE-AC02-05CH11231.

\end{acknowledgments}

\appendix*

\newpage
\section{Extraction of $\upeta/s$}

Below we explain our procedure for the extraction of the shear-viscosity to entropy density ratio $\upeta/s$ from the kinetic theory simulations. Starting point of the extraction, is the universal hydrodynamic late time behavior, where the ratio of longitudinal pressure to energy density behaves as 
\begin{equation}
\label{eq:appone}
    \frac{P_L}{\epsilon} = \frac{1}{3} - \frac{16}{9} \frac{\upeta/s}{\tau T} + \ldots .
\end{equation}
By inverting Eq.~(\ref{eq:appone}) for $\upeta/s$, we can use the ratio $9/16(1/3-P_L/\epsilon) T\tau$ to extract the ratio of shear viscosity to entropy density $\upeta/s$ from the asymptotic behavior of each numerical solution, and in subsequent analysis we use the mean value for each $\lambda$. We illustrate this extraction in the left panel of Fig.\,\ref{fig:viscosity_extraction}, while the right panel of Fig.\,\ref{fig:viscosity_extraction} shows the extracted values in comparison to NLL parametrization of $\upeta/s$ in \cite{Arnold:2003zc}. We note that for small couplings $\lambda$, the extracted values of $\upeta/s$ are in excellent agreement with the NLL parametrization, while for larger values of $\lambda$ the NLL parametrization breaks down, and as discussed in \cite{Kurkela:2018oqw} the transport coefficients also become more sensitive to the precise implementation of the screening of the elastic matrix elements. We also provide the extracted values in Table \ref{tab:viscosity}.

\vspace{0.3cm}
\begin{table}[ht]
\begin{tabular}{l|cccccccc}
\hline
$\lambda$ & 0.1 & 0.2 & 0.5 & 1 & 2 & 5 & 10 & 20 \\
\hline
$\upeta/s$ & 1361 & 394.9 & 79.69 & 24.55 & 7.845 & 1.848 & 0.6472 & 0.2313\\
\hline
\end{tabular}
\caption{Numerically extracted viscosity, averaged over different solutions. These values were used in the subsequent analysis.}
\label{tab:viscosity}
\end{table}

\newpage
\bibliography{ref.bib}

\begin{thebibliography}{76}%
\makeatletter
\providecommand \@ifxundefined [1]{%
 \@ifx{#1\undefined}
}%
\providecommand \@ifnum [1]{%
 \ifnum #1\expandafter \@firstoftwo
 \else \expandafter \@secondoftwo
 \fi
}%
\providecommand \@ifx [1]{%
 \ifx #1\expandafter \@firstoftwo
 \else \expandafter \@secondoftwo
 \fi
}%
\providecommand \natexlab [1]{#1}%
\providecommand \enquote  [1]{``#1''}%
\providecommand \bibnamefont  [1]{#1}%
\providecommand \bibfnamefont [1]{#1}%
\providecommand \citenamefont [1]{#1}%
\providecommand \href@noop [0]{\@secondoftwo}%
\providecommand \href [0]{\begingroup \@sanitize@url \@href}%
\providecommand \@href[1]{\@@startlink{#1}\@@href}%
\providecommand \@@href[1]{\endgroup#1\@@endlink}%
\providecommand \@sanitize@url [0]{\catcode `\\12\catcode `\$12\catcode
  `\&12\catcode `\#12\catcode `\^12\catcode `\_12\catcode `\%12\relax}%
\providecommand \@@startlink[1]{}%
\providecommand \@@endlink[0]{}%
\providecommand \url  [0]{\begingroup\@sanitize@url \@url }%
\providecommand \@url [1]{\endgroup\@href {#1}{\urlprefix }}%
\providecommand \urlprefix  [0]{URL }%
\providecommand \Eprint [0]{\href }%
\providecommand \doibase [0]{https://doi.org/}%
\providecommand \selectlanguage [0]{\@gobble}%
\providecommand \bibinfo  [0]{\@secondoftwo}%
\providecommand \bibfield  [0]{\@secondoftwo}%
\providecommand \translation [1]{[#1]}%
\providecommand \BibitemOpen [0]{}%
\providecommand \bibitemStop [0]{}%
\providecommand \bibitemNoStop [0]{.\EOS\space}%
\providecommand \EOS [0]{\spacefactor3000\relax}%
\providecommand \BibitemShut  [1]{\csname bibitem#1\endcsname}%
\let\auto@bib@innerbib\@empty
%</preamble>
\bibitem [{\citenamefont {Heinz}(2013)}]{Heinz:2013wva}%
  \BibitemOpen
  \bibfield  {author} {\bibinfo {author} {\bibfnamefont {U.~W.}\ \bibnamefont
  {Heinz}},\ }\bibfield  {title} {\bibinfo {title} {{Towards the Little Bang
  Standard Model}},\ }\href {https://doi.org/10.1088/1742-6596/455/1/012044}
  {\bibfield  {journal} {\bibinfo  {journal} {J. Phys. Conf. Ser.}\ }\textbf
  {\bibinfo {volume} {455}},\ \bibinfo {pages} {012044} (\bibinfo {year}
  {2013})},\ \Eprint {https://arxiv.org/abs/1304.3634} {arXiv:1304.3634
  [nucl-th]} \BibitemShut {NoStop}%
\bibitem [{\citenamefont {Schlichting}\ and\ \citenamefont
  {Teaney}(2019)}]{schlichting2019first}%
  \BibitemOpen
  \bibfield  {author} {\bibinfo {author} {\bibfnamefont {S.}~\bibnamefont
  {Schlichting}}\ and\ \bibinfo {author} {\bibfnamefont {D.}~\bibnamefont
  {Teaney}},\ }\bibfield  {title} {\bibinfo {title} {The first fm/c of
  heavy-ion collisions},\ }\href
  {https://doi.org/10.1146/annurev-nucl-101918-023825} {\bibfield  {journal}
  {\bibinfo  {journal} {Annual Review of Nuclear and Particle Science}\
  }\textbf {\bibinfo {volume} {69}},\ \bibinfo {pages} {447–476} (\bibinfo
  {year} {2019})}\BibitemShut {NoStop}%
\bibitem [{\citenamefont {Berges}\ \emph {et~al.}(2021)\citenamefont {Berges},
  \citenamefont {Heller}, \citenamefont {Mazeliauskas},\ and\ \citenamefont
  {Venugopalan}}]{Berges:2020thermalization}%
  \BibitemOpen
  \bibfield  {author} {\bibinfo {author} {\bibfnamefont {J.}~\bibnamefont
  {Berges}}, \bibinfo {author} {\bibfnamefont {M.~P.}\ \bibnamefont {Heller}},
  \bibinfo {author} {\bibfnamefont {A.}~\bibnamefont {Mazeliauskas}},\ and\
  \bibinfo {author} {\bibfnamefont {R.}~\bibnamefont {Venugopalan}},\
  }\bibfield  {title} {\bibinfo {title} {{QCD thermalization: Ab initio
  approaches and interdisciplinary connections}},\ }\href
  {https://doi.org/10.1103/RevModPhys.93.035003} {\bibfield  {journal}
  {\bibinfo  {journal} {Rev. Mod. Phys.}\ }\textbf {\bibinfo {volume} {93}},\
  \bibinfo {pages} {035003} (\bibinfo {year} {2021})},\ \Eprint
  {https://arxiv.org/abs/2005.12299} {arXiv:2005.12299 [hep-th]} \BibitemShut
  {NoStop}%
\bibitem [{\citenamefont
  {Romatschke}(2017{\natexlab{a}})}]{Romatschke:2016hle}%
  \BibitemOpen
  \bibfield  {author} {\bibinfo {author} {\bibfnamefont {P.}~\bibnamefont
  {Romatschke}},\ }\bibfield  {title} {\bibinfo {title} {{Do nuclear collisions
  create a locally equilibrated quark\textendash{}gluon plasma?}},\ }\href
  {https://doi.org/10.1140/epjc/s10052-016-4567-x} {\bibfield  {journal}
  {\bibinfo  {journal} {Eur. Phys. J. C}\ }\textbf {\bibinfo {volume} {77}},\
  \bibinfo {pages} {21} (\bibinfo {year} {2017}{\natexlab{a}})},\ \Eprint
  {https://arxiv.org/abs/1609.02820} {arXiv:1609.02820 [nucl-th]} \BibitemShut
  {NoStop}%
\bibitem [{\citenamefont {Florkowski}\ \emph {et~al.}(2018)\citenamefont
  {Florkowski}, \citenamefont {Heller},\ and\ \citenamefont
  {Spalinski}}]{Florkowski:2017olj}%
  \BibitemOpen
  \bibfield  {author} {\bibinfo {author} {\bibfnamefont {W.}~\bibnamefont
  {Florkowski}}, \bibinfo {author} {\bibfnamefont {M.~P.}\ \bibnamefont
  {Heller}},\ and\ \bibinfo {author} {\bibfnamefont {M.}~\bibnamefont
  {Spalinski}},\ }\bibfield  {title} {\bibinfo {title} {{New theories of
  relativistic hydrodynamics in the LHC era}},\ }\href
  {https://doi.org/10.1088/1361-6633/aaa091} {\bibfield  {journal} {\bibinfo
  {journal} {Rept. Prog. Phys.}\ }\textbf {\bibinfo {volume} {81}},\ \bibinfo
  {pages} {046001} (\bibinfo {year} {2018})},\ \Eprint
  {https://arxiv.org/abs/1707.02282} {arXiv:1707.02282 [hep-ph]} \BibitemShut
  {NoStop}%
\bibitem [{\citenamefont {Heller}\ and\ \citenamefont
  {Spalinski}(2015)}]{Heller:2015dha}%
  \BibitemOpen
  \bibfield  {author} {\bibinfo {author} {\bibfnamefont {M.~P.}\ \bibnamefont
  {Heller}}\ and\ \bibinfo {author} {\bibfnamefont {M.}~\bibnamefont
  {Spalinski}},\ }\bibfield  {title} {\bibinfo {title} {{Hydrodynamics Beyond
  the Gradient Expansion: Resurgence and Resummation}},\ }\href
  {https://doi.org/10.1103/PhysRevLett.115.072501} {\bibfield  {journal}
  {\bibinfo  {journal} {Phys. Rev. Lett.}\ }\textbf {\bibinfo {volume} {115}},\
  \bibinfo {pages} {072501} (\bibinfo {year} {2015})},\ \Eprint
  {https://arxiv.org/abs/1503.07514} {arXiv:1503.07514 [hep-th]} \BibitemShut
  {NoStop}%
\bibitem [{\citenamefont {Romatschke}(2018)}]{Romatschke:2017vte}%
  \BibitemOpen
  \bibfield  {author} {\bibinfo {author} {\bibfnamefont {P.}~\bibnamefont
  {Romatschke}},\ }\bibfield  {title} {\bibinfo {title} {{Relativistic Fluid
  Dynamics Far From Local Equilibrium}},\ }\href
  {https://doi.org/10.1103/PhysRevLett.120.012301} {\bibfield  {journal}
  {\bibinfo  {journal} {Phys. Rev. Lett.}\ }\textbf {\bibinfo {volume} {120}},\
  \bibinfo {pages} {012301} (\bibinfo {year} {2018})},\ \Eprint
  {https://arxiv.org/abs/1704.08699} {arXiv:1704.08699 [hep-th]} \BibitemShut
  {NoStop}%
\bibitem [{\citenamefont {Strickland}\ \emph {et~al.}(2018)\citenamefont
  {Strickland}, \citenamefont {Noronha},\ and\ \citenamefont
  {Denicol}}]{Strickland:2017kux}%
  \BibitemOpen
  \bibfield  {author} {\bibinfo {author} {\bibfnamefont {M.}~\bibnamefont
  {Strickland}}, \bibinfo {author} {\bibfnamefont {J.}~\bibnamefont
  {Noronha}},\ and\ \bibinfo {author} {\bibfnamefont {G.}~\bibnamefont
  {Denicol}},\ }\bibfield  {title} {\bibinfo {title} {{Anisotropic
  nonequilibrium hydrodynamic attractor}},\ }\href
  {https://doi.org/10.1103/PhysRevD.97.036020} {\bibfield  {journal} {\bibinfo
  {journal} {Phys. Rev. D}\ }\textbf {\bibinfo {volume} {97}},\ \bibinfo
  {pages} {036020} (\bibinfo {year} {2018})},\ \Eprint
  {https://arxiv.org/abs/1709.06644} {arXiv:1709.06644 [nucl-th]} \BibitemShut
  {NoStop}%
\bibitem [{\citenamefont {Strickland}(2018)}]{Strickland:2018ayk}%
  \BibitemOpen
  \bibfield  {author} {\bibinfo {author} {\bibfnamefont {M.}~\bibnamefont
  {Strickland}},\ }\bibfield  {title} {\bibinfo {title} {{The non-equilibrium
  attractor for kinetic theory in relaxation time approximation}},\ }\href
  {https://doi.org/10.1007/JHEP12(2018)128} {\bibfield  {journal} {\bibinfo
  {journal} {JHEP}\ }\textbf {\bibinfo {volume} {12}},\ \bibinfo {pages}
  {128}},\ \Eprint {https://arxiv.org/abs/1809.01200} {arXiv:1809.01200
  [nucl-th]} \BibitemShut {NoStop}%
\bibitem [{\citenamefont {Jaiswal}\ \emph {et~al.}(2019)\citenamefont
  {Jaiswal}, \citenamefont {Chattopadhyay}, \citenamefont {Jaiswal},
  \citenamefont {Pal},\ and\ \citenamefont {Heinz}}]{Jaiswal:2019cju}%
  \BibitemOpen
  \bibfield  {author} {\bibinfo {author} {\bibfnamefont {S.}~\bibnamefont
  {Jaiswal}}, \bibinfo {author} {\bibfnamefont {C.}~\bibnamefont
  {Chattopadhyay}}, \bibinfo {author} {\bibfnamefont {A.}~\bibnamefont
  {Jaiswal}}, \bibinfo {author} {\bibfnamefont {S.}~\bibnamefont {Pal}},\ and\
  \bibinfo {author} {\bibfnamefont {U.}~\bibnamefont {Heinz}},\ }\bibfield
  {title} {\bibinfo {title} {{Exact solutions and attractors of higher-order
  viscous fluid dynamics for Bjorken flow}},\ }\href
  {https://doi.org/10.1103/PhysRevC.100.034901} {\bibfield  {journal} {\bibinfo
   {journal} {Phys. Rev. C}\ }\textbf {\bibinfo {volume} {100}},\ \bibinfo
  {pages} {034901} (\bibinfo {year} {2019})},\ \Eprint
  {https://arxiv.org/abs/1907.07965} {arXiv:1907.07965 [nucl-th]} \BibitemShut
  {NoStop}%
\bibitem [{\citenamefont {Blaizot}\ and\ \citenamefont
  {Yan}(2020)}]{Blaizot:2019scw}%
  \BibitemOpen
  \bibfield  {author} {\bibinfo {author} {\bibfnamefont {J.-P.}\ \bibnamefont
  {Blaizot}}\ and\ \bibinfo {author} {\bibfnamefont {L.}~\bibnamefont {Yan}},\
  }\bibfield  {title} {\bibinfo {title} {{Emergence of hydrodynamical behavior
  in expanding ultra-relativistic plasmas}},\ }\href
  {https://doi.org/10.1016/j.aop.2019.167993} {\bibfield  {journal} {\bibinfo
  {journal} {Annals Phys.}\ }\textbf {\bibinfo {volume} {412}},\ \bibinfo
  {pages} {167993} (\bibinfo {year} {2020})},\ \Eprint
  {https://arxiv.org/abs/1904.08677} {arXiv:1904.08677 [nucl-th]} \BibitemShut
  {NoStop}%
\bibitem [{\citenamefont {Almaalol}\ \emph {et~al.}(2020)\citenamefont
  {Almaalol}, \citenamefont {Kurkela},\ and\ \citenamefont
  {Strickland}}]{Almaalol:2020rnu}%
  \BibitemOpen
  \bibfield  {author} {\bibinfo {author} {\bibfnamefont {D.}~\bibnamefont
  {Almaalol}}, \bibinfo {author} {\bibfnamefont {A.}~\bibnamefont {Kurkela}},\
  and\ \bibinfo {author} {\bibfnamefont {M.}~\bibnamefont {Strickland}},\
  }\bibfield  {title} {\bibinfo {title} {{Nonequilibrium Attractor in
  High-Temperature QCD Plasmas}},\ }\href
  {https://doi.org/10.1103/PhysRevLett.125.122302} {\bibfield  {journal}
  {\bibinfo  {journal} {Phys. Rev. Lett.}\ }\textbf {\bibinfo {volume} {125}},\
  \bibinfo {pages} {122302} (\bibinfo {year} {2020})},\ \Eprint
  {https://arxiv.org/abs/2004.05195} {arXiv:2004.05195 [hep-ph]} \BibitemShut
  {NoStop}%
\bibitem [{\citenamefont {Heller}\ \emph {et~al.}(2020)\citenamefont {Heller},
  \citenamefont {Jefferson}, \citenamefont {Spali\'nski},\ and\ \citenamefont
  {Svensson}}]{Heller:2020anv}%
  \BibitemOpen
  \bibfield  {author} {\bibinfo {author} {\bibfnamefont {M.~P.}\ \bibnamefont
  {Heller}}, \bibinfo {author} {\bibfnamefont {R.}~\bibnamefont {Jefferson}},
  \bibinfo {author} {\bibfnamefont {M.}~\bibnamefont {Spali\'nski}},\ and\
  \bibinfo {author} {\bibfnamefont {V.}~\bibnamefont {Svensson}},\ }\bibfield
  {title} {\bibinfo {title} {{Hydrodynamic Attractors in Phase Space}},\ }\href
  {https://doi.org/10.1103/PhysRevLett.125.132301} {\bibfield  {journal}
  {\bibinfo  {journal} {Phys. Rev. Lett.}\ }\textbf {\bibinfo {volume} {125}},\
  \bibinfo {pages} {132301} (\bibinfo {year} {2020})},\ \Eprint
  {https://arxiv.org/abs/2003.07368} {arXiv:2003.07368 [hep-th]} \BibitemShut
  {NoStop}%
\bibitem [{\citenamefont {Du}\ and\ \citenamefont
  {Schlichting}(2021{\natexlab{a}})}]{Du:2020zqg}%
  \BibitemOpen
  \bibfield  {author} {\bibinfo {author} {\bibfnamefont {X.}~\bibnamefont
  {Du}}\ and\ \bibinfo {author} {\bibfnamefont {S.}~\bibnamefont
  {Schlichting}},\ }\bibfield  {title} {\bibinfo {title} {{Equilibration of the
  Quark-Gluon Plasma at Finite Net-Baryon Density in QCD Kinetic Theory}},\
  }\href {https://doi.org/10.1103/PhysRevLett.127.122301} {\bibfield  {journal}
  {\bibinfo  {journal} {Phys. Rev. Lett.}\ }\textbf {\bibinfo {volume} {127}},\
  \bibinfo {pages} {122301} (\bibinfo {year} {2021}{\natexlab{a}})},\ \Eprint
  {https://arxiv.org/abs/2012.09068} {arXiv:2012.09068 [hep-ph]} \BibitemShut
  {NoStop}%
\bibitem [{\citenamefont
  {Romatschke}(2017{\natexlab{b}})}]{Romatschke:2017acs}%
  \BibitemOpen
  \bibfield  {author} {\bibinfo {author} {\bibfnamefont {P.}~\bibnamefont
  {Romatschke}},\ }\bibfield  {title} {\bibinfo {title} {{Relativistic
  Hydrodynamic Attractors with Broken Symmetries: Non-Conformal and
  Non-Homogeneous}},\ }\href {https://doi.org/10.1007/JHEP12(2017)079}
  {\bibfield  {journal} {\bibinfo  {journal} {JHEP}\ }\textbf {\bibinfo
  {volume} {12}},\ \bibinfo {pages} {079}},\ \Eprint
  {https://arxiv.org/abs/1710.03234} {arXiv:1710.03234 [hep-th]} \BibitemShut
  {NoStop}%
\bibitem [{\citenamefont {Behtash}\ \emph {et~al.}(2018)\citenamefont
  {Behtash}, \citenamefont {Cruz-Camacho},\ and\ \citenamefont
  {Martinez}}]{Behtash:2017wqg}%
  \BibitemOpen
  \bibfield  {author} {\bibinfo {author} {\bibfnamefont {A.}~\bibnamefont
  {Behtash}}, \bibinfo {author} {\bibfnamefont {C.~N.}\ \bibnamefont
  {Cruz-Camacho}},\ and\ \bibinfo {author} {\bibfnamefont {M.}~\bibnamefont
  {Martinez}},\ }\bibfield  {title} {\bibinfo {title} {{Far-from-equilibrium
  attractors and nonlinear dynamical systems approach to the Gubser flow}},\
  }\href {https://doi.org/10.1103/PhysRevD.97.044041} {\bibfield  {journal}
  {\bibinfo  {journal} {Phys. Rev. D}\ }\textbf {\bibinfo {volume} {97}},\
  \bibinfo {pages} {044041} (\bibinfo {year} {2018})},\ \Eprint
  {https://arxiv.org/abs/1711.01745} {arXiv:1711.01745 [hep-th]} \BibitemShut
  {NoStop}%
\bibitem [{\citenamefont {Denicol}\ and\ \citenamefont
  {Noronha}(2018)}]{Denicol:2017lxn}%
  \BibitemOpen
  \bibfield  {author} {\bibinfo {author} {\bibfnamefont {G.~S.}\ \bibnamefont
  {Denicol}}\ and\ \bibinfo {author} {\bibfnamefont {J.}~\bibnamefont
  {Noronha}},\ }\bibfield  {title} {\bibinfo {title} {{Analytical attractor and
  the divergence of the slow-roll expansion in relativistic hydrodynamics}},\
  }\href {https://doi.org/10.1103/PhysRevD.97.056021} {\bibfield  {journal}
  {\bibinfo  {journal} {Phys. Rev. D}\ }\textbf {\bibinfo {volume} {97}},\
  \bibinfo {pages} {056021} (\bibinfo {year} {2018})},\ \Eprint
  {https://arxiv.org/abs/1711.01657} {arXiv:1711.01657 [nucl-th]} \BibitemShut
  {NoStop}%
\bibitem [{\citenamefont {Denicol}\ and\ \citenamefont
  {Noronha}(2020)}]{Denicol:2019lio}%
  \BibitemOpen
  \bibfield  {author} {\bibinfo {author} {\bibfnamefont {G.~S.}\ \bibnamefont
  {Denicol}}\ and\ \bibinfo {author} {\bibfnamefont {J.}~\bibnamefont
  {Noronha}},\ }\bibfield  {title} {\bibinfo {title} {{Exact hydrodynamic
  attractor of an ultrarelativistic gas of hard spheres}},\ }\href
  {https://doi.org/10.1103/PhysRevLett.124.152301} {\bibfield  {journal}
  {\bibinfo  {journal} {Phys. Rev. Lett.}\ }\textbf {\bibinfo {volume} {124}},\
  \bibinfo {pages} {152301} (\bibinfo {year} {2020})},\ \Eprint
  {https://arxiv.org/abs/1908.09957} {arXiv:1908.09957 [nucl-th]} \BibitemShut
  {NoStop}%
\bibitem [{\citenamefont {Kurkela}\ \emph {et~al.}(2020)\citenamefont
  {Kurkela}, \citenamefont {van~der Schee}, \citenamefont {Wiedemann},\ and\
  \citenamefont {Wu}}]{Kurkela:2019set}%
  \BibitemOpen
  \bibfield  {author} {\bibinfo {author} {\bibfnamefont {A.}~\bibnamefont
  {Kurkela}}, \bibinfo {author} {\bibfnamefont {W.}~\bibnamefont {van~der
  Schee}}, \bibinfo {author} {\bibfnamefont {U.~A.}\ \bibnamefont
  {Wiedemann}},\ and\ \bibinfo {author} {\bibfnamefont {B.}~\bibnamefont
  {Wu}},\ }\bibfield  {title} {\bibinfo {title} {{Early- and Late-Time Behavior
  of Attractors in Heavy-Ion Collisions}},\ }\href
  {https://doi.org/10.1103/PhysRevLett.124.102301} {\bibfield  {journal}
  {\bibinfo  {journal} {Phys. Rev. Lett.}\ }\textbf {\bibinfo {volume} {124}},\
  \bibinfo {pages} {102301} (\bibinfo {year} {2020})},\ \Eprint
  {https://arxiv.org/abs/1907.08101} {arXiv:1907.08101 [hep-ph]} \BibitemShut
  {NoStop}%
\bibitem [{\citenamefont {Blaizot}\ and\ \citenamefont
  {Yan}(2021)}]{Blaizot:2020gql}%
  \BibitemOpen
  \bibfield  {author} {\bibinfo {author} {\bibfnamefont {J.-P.}\ \bibnamefont
  {Blaizot}}\ and\ \bibinfo {author} {\bibfnamefont {L.}~\bibnamefont {Yan}},\
  }\bibfield  {title} {\bibinfo {title} {{Analytical attractor for Bjorken
  flows}},\ }\href {https://doi.org/10.1016/j.physletb.2021.136478} {\bibfield
  {journal} {\bibinfo  {journal} {Phys. Lett. B}\ }\textbf {\bibinfo {volume}
  {820}},\ \bibinfo {pages} {136478} (\bibinfo {year} {2021})},\ \Eprint
  {https://arxiv.org/abs/2006.08815} {arXiv:2006.08815 [nucl-th]} \BibitemShut
  {NoStop}%
\bibitem [{\citenamefont {Behtash}\ \emph {et~al.}(2021)\citenamefont
  {Behtash}, \citenamefont {Kamata}, \citenamefont {Martinez}, \citenamefont
  {Sch\"afer},\ and\ \citenamefont {Skokov}}]{Behtash:2020vqk}%
  \BibitemOpen
  \bibfield  {author} {\bibinfo {author} {\bibfnamefont {A.}~\bibnamefont
  {Behtash}}, \bibinfo {author} {\bibfnamefont {S.}~\bibnamefont {Kamata}},
  \bibinfo {author} {\bibfnamefont {M.}~\bibnamefont {Martinez}}, \bibinfo
  {author} {\bibfnamefont {T.}~\bibnamefont {Sch\"afer}},\ and\ \bibinfo
  {author} {\bibfnamefont {V.}~\bibnamefont {Skokov}},\ }\bibfield  {title}
  {\bibinfo {title} {{Transasymptotics and hydrodynamization of the
  Fokker-Planck equation for gluons}},\ }\href
  {https://doi.org/10.1103/PhysRevD.103.056010} {\bibfield  {journal} {\bibinfo
   {journal} {Phys. Rev. D}\ }\textbf {\bibinfo {volume} {103}},\ \bibinfo
  {pages} {056010} (\bibinfo {year} {2021})},\ \Eprint
  {https://arxiv.org/abs/2011.08235} {arXiv:2011.08235 [hep-ph]} \BibitemShut
  {NoStop}%
\bibitem [{\citenamefont {Du}\ \emph {et~al.}(2021)\citenamefont {Du},
  \citenamefont {Huang},\ and\ \citenamefont {Taya}}]{Du:2021fok}%
  \BibitemOpen
  \bibfield  {author} {\bibinfo {author} {\bibfnamefont {Z.}~\bibnamefont
  {Du}}, \bibinfo {author} {\bibfnamefont {X.-G.}\ \bibnamefont {Huang}},\ and\
  \bibinfo {author} {\bibfnamefont {H.}~\bibnamefont {Taya}},\ }\bibfield
  {title} {\bibinfo {title} {{Hydrodynamic attractor in a Hubble expansion}},\
  }\href {https://doi.org/10.1103/PhysRevD.104.056022} {\bibfield  {journal}
  {\bibinfo  {journal} {Phys. Rev. D}\ }\textbf {\bibinfo {volume} {104}},\
  \bibinfo {pages} {056022} (\bibinfo {year} {2021})},\ \Eprint
  {https://arxiv.org/abs/2104.12534} {arXiv:2104.12534 [nucl-th]} \BibitemShut
  {NoStop}%
\bibitem [{\citenamefont {Chattopadhyay}\ \emph {et~al.}(2022)\citenamefont
  {Chattopadhyay}, \citenamefont {Jaiswal}, \citenamefont {Du}, \citenamefont
  {Heinz},\ and\ \citenamefont {Pal}}]{Chattopadhyay:2021ive}%
  \BibitemOpen
  \bibfield  {author} {\bibinfo {author} {\bibfnamefont {C.}~\bibnamefont
  {Chattopadhyay}}, \bibinfo {author} {\bibfnamefont {S.}~\bibnamefont
  {Jaiswal}}, \bibinfo {author} {\bibfnamefont {L.}~\bibnamefont {Du}},
  \bibinfo {author} {\bibfnamefont {U.}~\bibnamefont {Heinz}},\ and\ \bibinfo
  {author} {\bibfnamefont {S.}~\bibnamefont {Pal}},\ }\bibfield  {title}
  {\bibinfo {title} {{Non-conformal attractor in boost-invariant plasmas}},\
  }\href {https://doi.org/10.1016/j.physletb.2021.136820} {\bibfield  {journal}
  {\bibinfo  {journal} {Phys. Lett. B}\ }\textbf {\bibinfo {volume} {824}},\
  \bibinfo {pages} {136820} (\bibinfo {year} {2022})},\ \Eprint
  {https://arxiv.org/abs/2107.05500} {arXiv:2107.05500 [nucl-th]} \BibitemShut
  {NoStop}%
\bibitem [{\citenamefont {Kurkela}\ \emph
  {et~al.}(2019{\natexlab{a}})\citenamefont {Kurkela}, \citenamefont
  {Mazeliauskas}, \citenamefont {Paquet}, \citenamefont {Schlichting},\ and\
  \citenamefont {Teaney}}]{Kurkela:2018wud}%
  \BibitemOpen
  \bibfield  {author} {\bibinfo {author} {\bibfnamefont {A.}~\bibnamefont
  {Kurkela}}, \bibinfo {author} {\bibfnamefont {A.}~\bibnamefont
  {Mazeliauskas}}, \bibinfo {author} {\bibfnamefont {J.-F.}\ \bibnamefont
  {Paquet}}, \bibinfo {author} {\bibfnamefont {S.}~\bibnamefont
  {Schlichting}},\ and\ \bibinfo {author} {\bibfnamefont {D.}~\bibnamefont
  {Teaney}},\ }\bibfield  {title} {\bibinfo {title} {{Matching the
  Nonequilibrium Initial Stage of Heavy Ion Collisions to Hydrodynamics with
  QCD Kinetic Theory}},\ }\href
  {https://doi.org/10.1103/PhysRevLett.122.122302} {\bibfield  {journal}
  {\bibinfo  {journal} {Phys. Rev. Lett.}\ }\textbf {\bibinfo {volume} {122}},\
  \bibinfo {pages} {122302} (\bibinfo {year} {2019}{\natexlab{a}})},\ \Eprint
  {https://arxiv.org/abs/1805.01604} {arXiv:1805.01604 [hep-ph]} \BibitemShut
  {NoStop}%
\bibitem [{\citenamefont {Giacalone}\ \emph {et~al.}(2019)\citenamefont
  {Giacalone}, \citenamefont {Mazeliauskas},\ and\ \citenamefont
  {Schlichting}}]{Giacalone:2019ldn}%
  \BibitemOpen
  \bibfield  {author} {\bibinfo {author} {\bibfnamefont {G.}~\bibnamefont
  {Giacalone}}, \bibinfo {author} {\bibfnamefont {A.}~\bibnamefont
  {Mazeliauskas}},\ and\ \bibinfo {author} {\bibfnamefont {S.}~\bibnamefont
  {Schlichting}},\ }\bibfield  {title} {\bibinfo {title} {{Hydrodynamic
  attractors, initial state energy and particle production in relativistic
  nuclear collisions}},\ }\href
  {https://doi.org/10.1103/PhysRevLett.123.262301} {\bibfield  {journal}
  {\bibinfo  {journal} {Phys. Rev. Lett.}\ }\textbf {\bibinfo {volume} {123}},\
  \bibinfo {pages} {262301} (\bibinfo {year} {2019})},\ \Eprint
  {https://arxiv.org/abs/1908.02866} {arXiv:1908.02866 [hep-ph]} \BibitemShut
  {NoStop}%
\bibitem [{\citenamefont {Jankowski}\ \emph {et~al.}(2021)\citenamefont
  {Jankowski}, \citenamefont {Kamata}, \citenamefont {Martinez},\ and\
  \citenamefont {Spali\'nski}}]{Jankowski:2020itt}%
  \BibitemOpen
  \bibfield  {author} {\bibinfo {author} {\bibfnamefont {J.}~\bibnamefont
  {Jankowski}}, \bibinfo {author} {\bibfnamefont {S.}~\bibnamefont {Kamata}},
  \bibinfo {author} {\bibfnamefont {M.}~\bibnamefont {Martinez}},\ and\
  \bibinfo {author} {\bibfnamefont {M.}~\bibnamefont {Spali\'nski}},\
  }\bibfield  {title} {\bibinfo {title} {{Constraining the initial stages of
  ultrarelativistic nuclear collisions}},\ }\href
  {https://doi.org/10.1103/PhysRevD.104.074012} {\bibfield  {journal} {\bibinfo
   {journal} {Phys. Rev. D}\ }\textbf {\bibinfo {volume} {104}},\ \bibinfo
  {pages} {074012} (\bibinfo {year} {2021})},\ \Eprint
  {https://arxiv.org/abs/2012.02184} {arXiv:2012.02184 [nucl-th]} \BibitemShut
  {NoStop}%
\bibitem [{\citenamefont {Coquet}\ \emph {et~al.}(2021)\citenamefont {Coquet},
  \citenamefont {Du}, \citenamefont {Ollitrault}, \citenamefont {Schlichting},\
  and\ \citenamefont {Winn}}]{Coquet:2021lca}%
  \BibitemOpen
  \bibfield  {author} {\bibinfo {author} {\bibfnamefont {M.}~\bibnamefont
  {Coquet}}, \bibinfo {author} {\bibfnamefont {X.}~\bibnamefont {Du}}, \bibinfo
  {author} {\bibfnamefont {J.-Y.}\ \bibnamefont {Ollitrault}}, \bibinfo
  {author} {\bibfnamefont {S.}~\bibnamefont {Schlichting}},\ and\ \bibinfo
  {author} {\bibfnamefont {M.}~\bibnamefont {Winn}},\ }\bibfield  {title}
  {\bibinfo {title} {{Intermediate mass dileptons as pre-equilibrium probes in
  heavy ion collisions}},\ }\href
  {https://doi.org/10.1016/j.physletb.2021.136626} {\bibfield  {journal}
  {\bibinfo  {journal} {Phys. Lett. B}\ }\textbf {\bibinfo {volume} {821}},\
  \bibinfo {pages} {136626} (\bibinfo {year} {2021})},\ \Eprint
  {https://arxiv.org/abs/2104.07622} {arXiv:2104.07622 [nucl-th]} \BibitemShut
  {NoStop}%
\bibitem [{\citenamefont {Soloviev}(2021)}]{Soloviev:2021lhs}%
  \BibitemOpen
  \bibfield  {author} {\bibinfo {author} {\bibfnamefont {A.}~\bibnamefont
  {Soloviev}},\ }\bibfield  {title} {\bibinfo {title} {{Hydrodynamic attractors
  in heavy ion collisions: a review}},\ }\href@noop {} {\  (\bibinfo {year}
  {2021})},\ \Eprint {https://arxiv.org/abs/2109.15081} {arXiv:2109.15081
  [hep-th]} \BibitemShut {NoStop}%
\bibitem [{\citenamefont {Ambrus}\ \emph {et~al.}(2021)\citenamefont {Ambrus},
  \citenamefont {Busuioc}, \citenamefont {Fotakis}, \citenamefont
  {Gallmeister},\ and\ \citenamefont {Greiner}}]{Ambrus:2021sjg}%
  \BibitemOpen
  \bibfield  {author} {\bibinfo {author} {\bibfnamefont {V.~E.}\ \bibnamefont
  {Ambrus}}, \bibinfo {author} {\bibfnamefont {S.}~\bibnamefont {Busuioc}},
  \bibinfo {author} {\bibfnamefont {J.~A.}\ \bibnamefont {Fotakis}}, \bibinfo
  {author} {\bibfnamefont {K.}~\bibnamefont {Gallmeister}},\ and\ \bibinfo
  {author} {\bibfnamefont {C.}~\bibnamefont {Greiner}},\ }\bibfield  {title}
  {\bibinfo {title} {{Bjorken flow attractors with transverse dynamics}},\
  }\href {https://doi.org/10.1103/PhysRevD.104.094022} {\bibfield  {journal}
  {\bibinfo  {journal} {Phys. Rev. D}\ }\textbf {\bibinfo {volume} {104}},\
  \bibinfo {pages} {094022} (\bibinfo {year} {2021})},\ \Eprint
  {https://arxiv.org/abs/2102.11785} {arXiv:2102.11785 [nucl-th]} \BibitemShut
  {NoStop}%
\bibitem [{\citenamefont {Gelis}\ \emph {et~al.}(2010)\citenamefont {Gelis},
  \citenamefont {Iancu}, \citenamefont {Jalilian-Marian},\ and\ \citenamefont
  {Venugopalan}}]{Gelis:2010nm}%
  \BibitemOpen
  \bibfield  {author} {\bibinfo {author} {\bibfnamefont {F.}~\bibnamefont
  {Gelis}}, \bibinfo {author} {\bibfnamefont {E.}~\bibnamefont {Iancu}},
  \bibinfo {author} {\bibfnamefont {J.}~\bibnamefont {Jalilian-Marian}},\ and\
  \bibinfo {author} {\bibfnamefont {R.}~\bibnamefont {Venugopalan}},\
  }\bibfield  {title} {\bibinfo {title} {{The Color Glass Condensate}},\ }\href
  {https://doi.org/10.1146/annurev.nucl.010909.083629} {\bibfield  {journal}
  {\bibinfo  {journal} {Ann. Rev. Nucl. Part. Sci.}\ }\textbf {\bibinfo
  {volume} {60}},\ \bibinfo {pages} {463} (\bibinfo {year} {2010})},\ \Eprint
  {https://arxiv.org/abs/1002.0333} {arXiv:1002.0333 [hep-ph]} \BibitemShut
  {NoStop}%
\bibitem [{\citenamefont {Iancu}\ and\ \citenamefont
  {Venugopalan}(2003)}]{Iancu:2003xm}%
  \BibitemOpen
  \bibfield  {author} {\bibinfo {author} {\bibfnamefont {E.}~\bibnamefont
  {Iancu}}\ and\ \bibinfo {author} {\bibfnamefont {R.}~\bibnamefont
  {Venugopalan}},\ }\bibinfo {title} {{The Color glass condensate and
  high-energy scattering in QCD}},\ in\ \href
  {https://doi.org/10.1142/9789812795533_0005} {\emph {\bibinfo {booktitle}
  {{Quark-gluon plasma 4}}}},\ \bibinfo {editor} {edited by\ \bibinfo {editor}
  {\bibfnamefont {R.~C.}\ \bibnamefont {Hwa}}\ and\ \bibinfo {editor}
  {\bibfnamefont {X.-N.}\ \bibnamefont {Wang}}}\ (\bibinfo {year} {2003})\ pp.\
  \bibinfo {pages} {249--3363},\ \Eprint {https://arxiv.org/abs/hep-ph/0303204}
  {arXiv:hep-ph/0303204} \BibitemShut {NoStop}%
\bibitem [{\citenamefont {Blaizot}\ and\ \citenamefont
  {Yan}(2018)}]{Blaizot:2017ucy}%
  \BibitemOpen
  \bibfield  {author} {\bibinfo {author} {\bibfnamefont {J.-P.}\ \bibnamefont
  {Blaizot}}\ and\ \bibinfo {author} {\bibfnamefont {L.}~\bibnamefont {Yan}},\
  }\bibfield  {title} {\bibinfo {title} {{Fluid dynamics of out of equilibrium
  boost invariant plasmas}},\ }\href
  {https://doi.org/10.1016/j.physletb.2018.02.058} {\bibfield  {journal}
  {\bibinfo  {journal} {Phys. Lett. B}\ }\textbf {\bibinfo {volume} {780}},\
  \bibinfo {pages} {283} (\bibinfo {year} {2018})},\ \Eprint
  {https://arxiv.org/abs/1712.03856} {arXiv:1712.03856 [nucl-th]} \BibitemShut
  {NoStop}%
\bibitem [{\citenamefont {Heller}\ \emph
  {et~al.}(2018{\natexlab{a}})\citenamefont {Heller}, \citenamefont {Kurkela},
  \citenamefont {Spali{\'n}ski},\ and\ \citenamefont
  {Svensson}}]{hellerHydrodynamizationKineticTheory2018}%
  \BibitemOpen
  \bibfield  {author} {\bibinfo {author} {\bibfnamefont {M.~P.}\ \bibnamefont
  {Heller}}, \bibinfo {author} {\bibfnamefont {A.}~\bibnamefont {Kurkela}},
  \bibinfo {author} {\bibfnamefont {M.}~\bibnamefont {Spali{\'n}ski}},\ and\
  \bibinfo {author} {\bibfnamefont {V.}~\bibnamefont {Svensson}},\ }\bibfield
  {title} {\bibinfo {title} {Hydrodynamization in kinetic theory: {{Transient}}
  modes and the gradient expansion},\ }\href
  {https://doi.org/10.1103/PhysRevD.97.091503} {\bibfield  {journal} {\bibinfo
  {journal} {Physical Review D}\ }\textbf {\bibinfo {volume} {97}},\ \bibinfo
  {pages} {091503} (\bibinfo {year} {2018}{\natexlab{a}})}\BibitemShut
  {NoStop}%
\bibitem [{\citenamefont {Heller}\ and\ \citenamefont
  {Svensson}(2018)}]{hellerHowDoesRelativistic2018}%
  \BibitemOpen
  \bibfield  {author} {\bibinfo {author} {\bibfnamefont {M.~P.}\ \bibnamefont
  {Heller}}\ and\ \bibinfo {author} {\bibfnamefont {V.}~\bibnamefont
  {Svensson}},\ }\bibfield  {title} {\bibinfo {title} {How does relativistic
  kinetic theory remember about initial conditions?},\ }\href
  {https://doi.org/10.1103/PhysRevD.98.054016} {\bibfield  {journal} {\bibinfo
  {journal} {Physical Review D}\ }\textbf {\bibinfo {volume} {98}},\ \bibinfo
  {pages} {054016} (\bibinfo {year} {2018})}\BibitemShut {NoStop}%
\bibitem [{\citenamefont {Blaizot}\ \emph {et~al.}(2010)\citenamefont
  {Blaizot}, \citenamefont {Lappi},\ and\ \citenamefont
  {Mehtar-Tani}}]{Blaizot:2010kh}%
  \BibitemOpen
  \bibfield  {author} {\bibinfo {author} {\bibfnamefont {J.~P.}\ \bibnamefont
  {Blaizot}}, \bibinfo {author} {\bibfnamefont {T.}~\bibnamefont {Lappi}},\
  and\ \bibinfo {author} {\bibfnamefont {Y.}~\bibnamefont {Mehtar-Tani}},\
  }\bibfield  {title} {\bibinfo {title} {{On the gluon spectrum in the
  glasma}},\ }\href {https://doi.org/10.1016/j.nuclphysa.2010.06.009}
  {\bibfield  {journal} {\bibinfo  {journal} {Nucl. Phys. A}\ }\textbf
  {\bibinfo {volume} {846}},\ \bibinfo {pages} {63} (\bibinfo {year} {2010})},\
  \Eprint {https://arxiv.org/abs/1005.0955} {arXiv:1005.0955 [hep-ph]}
  \BibitemShut {NoStop}%
\bibitem [{\citenamefont {Lappi}\ and\ \citenamefont
  {Schlichting}(2018)}]{Lappi:2017skr}%
  \BibitemOpen
  \bibfield  {author} {\bibinfo {author} {\bibfnamefont {T.}~\bibnamefont
  {Lappi}}\ and\ \bibinfo {author} {\bibfnamefont {S.}~\bibnamefont
  {Schlichting}},\ }\bibfield  {title} {\bibinfo {title} {{Linearly polarized
  gluons and axial charge fluctuations in the Glasma}},\ }\href
  {https://doi.org/10.1103/PhysRevD.97.034034} {\bibfield  {journal} {\bibinfo
  {journal} {Phys. Rev. D}\ }\textbf {\bibinfo {volume} {97}},\ \bibinfo
  {pages} {034034} (\bibinfo {year} {2018})},\ \Eprint
  {https://arxiv.org/abs/1708.08625} {arXiv:1708.08625 [hep-ph]} \BibitemShut
  {NoStop}%
\bibitem [{\citenamefont {Golec-Biernat}\ and\ \citenamefont
  {Wusthoff}(1998)}]{Golec-Biernat:1998zce}%
  \BibitemOpen
  \bibfield  {author} {\bibinfo {author} {\bibfnamefont {K.~J.}\ \bibnamefont
  {Golec-Biernat}}\ and\ \bibinfo {author} {\bibfnamefont {M.}~\bibnamefont
  {Wusthoff}},\ }\bibfield  {title} {\bibinfo {title} {{Saturation effects in
  deep inelastic scattering at low Q**2 and its implications on diffraction}},\
  }\href {https://doi.org/10.1103/PhysRevD.59.014017} {\bibfield  {journal}
  {\bibinfo  {journal} {Phys. Rev. D}\ }\textbf {\bibinfo {volume} {59}},\
  \bibinfo {pages} {014017} (\bibinfo {year} {1998})},\ \Eprint
  {https://arxiv.org/abs/hep-ph/9807513} {arXiv:hep-ph/9807513} \BibitemShut
  {NoStop}%
\bibitem [{\citenamefont {Albacete}\ and\ \citenamefont
  {Marquet}(2014)}]{Albacete:2014fwa}%
  \BibitemOpen
  \bibfield  {author} {\bibinfo {author} {\bibfnamefont {J.~L.}\ \bibnamefont
  {Albacete}}\ and\ \bibinfo {author} {\bibfnamefont {C.}~\bibnamefont
  {Marquet}},\ }\bibfield  {title} {\bibinfo {title} {{Gluon saturation and
  initial conditions for relativistic heavy ion collisions}},\ }\href
  {https://doi.org/10.1016/j.ppnp.2014.01.004} {\bibfield  {journal} {\bibinfo
  {journal} {Prog. Part. Nucl. Phys.}\ }\textbf {\bibinfo {volume} {76}},\
  \bibinfo {pages} {1} (\bibinfo {year} {2014})},\ \Eprint
  {https://arxiv.org/abs/1401.4866} {arXiv:1401.4866 [hep-ph]} \BibitemShut
  {NoStop}%
\bibitem [{\citenamefont {Dumitru}\ and\ \citenamefont
  {McLerran}(2002)}]{Dumitru:2001ux}%
  \BibitemOpen
  \bibfield  {author} {\bibinfo {author} {\bibfnamefont {A.}~\bibnamefont
  {Dumitru}}\ and\ \bibinfo {author} {\bibfnamefont {L.~D.}\ \bibnamefont
  {McLerran}},\ }\bibfield  {title} {\bibinfo {title} {{How protons shatter
  colored glass}},\ }\href {https://doi.org/10.1016/S0375-9474(01)01301-X}
  {\bibfield  {journal} {\bibinfo  {journal} {Nucl. Phys. A}\ }\textbf
  {\bibinfo {volume} {700}},\ \bibinfo {pages} {492} (\bibinfo {year}
  {2002})},\ \Eprint {https://arxiv.org/abs/hep-ph/0105268}
  {arXiv:hep-ph/0105268} \BibitemShut {NoStop}%
\bibitem [{\citenamefont {Greif}\ \emph {et~al.}(2017)\citenamefont {Greif},
  \citenamefont {Greiner}, \citenamefont {Schenke}, \citenamefont
  {Schlichting},\ and\ \citenamefont {Xu}}]{Greif:2017bnr}%
  \BibitemOpen
  \bibfield  {author} {\bibinfo {author} {\bibfnamefont {M.}~\bibnamefont
  {Greif}}, \bibinfo {author} {\bibfnamefont {C.}~\bibnamefont {Greiner}},
  \bibinfo {author} {\bibfnamefont {B.}~\bibnamefont {Schenke}}, \bibinfo
  {author} {\bibfnamefont {S.}~\bibnamefont {Schlichting}},\ and\ \bibinfo
  {author} {\bibfnamefont {Z.}~\bibnamefont {Xu}},\ }\bibfield  {title}
  {\bibinfo {title} {{Importance of initial and final state effects for
  azimuthal correlations in p+Pb collisions}},\ }\href
  {https://doi.org/10.1103/PhysRevD.96.091504} {\bibfield  {journal} {\bibinfo
  {journal} {Phys. Rev. D}\ }\textbf {\bibinfo {volume} {96}},\ \bibinfo
  {pages} {091504} (\bibinfo {year} {2017})},\ \Eprint
  {https://arxiv.org/abs/1708.02076} {arXiv:1708.02076 [hep-ph]} \BibitemShut
  {NoStop}%
\bibitem [{\citenamefont {Kurkela}\ and\ \citenamefont
  {Zhu}(2015)}]{Kurkela:2015qoa}%
  \BibitemOpen
  \bibfield  {author} {\bibinfo {author} {\bibfnamefont {A.}~\bibnamefont
  {Kurkela}}\ and\ \bibinfo {author} {\bibfnamefont {Y.}~\bibnamefont {Zhu}},\
  }\bibfield  {title} {\bibinfo {title} {{Isotropization and hydrodynamization
  in weakly coupled heavy-ion collisions}},\ }\href
  {https://doi.org/10.1103/PhysRevLett.115.182301} {\bibfield  {journal}
  {\bibinfo  {journal} {Phys. Rev. Lett.}\ }\textbf {\bibinfo {volume} {115}},\
  \bibinfo {pages} {182301} (\bibinfo {year} {2015})},\ \Eprint
  {https://arxiv.org/abs/1506.06647} {arXiv:1506.06647 [hep-ph]} \BibitemShut
  {NoStop}%
\bibitem [{\citenamefont {Kurkela}\ \emph
  {et~al.}(2019{\natexlab{b}})\citenamefont {Kurkela}, \citenamefont
  {Mazeliauskas}, \citenamefont {Paquet}, \citenamefont {Schlichting},\ and\
  \citenamefont {Teaney}}]{Kurkela:2018vqr}%
  \BibitemOpen
  \bibfield  {author} {\bibinfo {author} {\bibfnamefont {A.}~\bibnamefont
  {Kurkela}}, \bibinfo {author} {\bibfnamefont {A.}~\bibnamefont
  {Mazeliauskas}}, \bibinfo {author} {\bibfnamefont {J.-F.}\ \bibnamefont
  {Paquet}}, \bibinfo {author} {\bibfnamefont {S.}~\bibnamefont
  {Schlichting}},\ and\ \bibinfo {author} {\bibfnamefont {D.}~\bibnamefont
  {Teaney}},\ }\bibfield  {title} {\bibinfo {title} {{Effective kinetic
  description of event-by-event pre-equilibrium dynamics in high-energy
  heavy-ion collisions}},\ }\href {https://doi.org/10.1103/PhysRevC.99.034910}
  {\bibfield  {journal} {\bibinfo  {journal} {Phys. Rev. C}\ }\textbf {\bibinfo
  {volume} {99}},\ \bibinfo {pages} {034910} (\bibinfo {year}
  {2019}{\natexlab{b}})},\ \Eprint {https://arxiv.org/abs/1805.00961}
  {arXiv:1805.00961 [hep-ph]} \BibitemShut {NoStop}%
\bibitem [{\citenamefont {Arnold}\ \emph
  {et~al.}(2003{\natexlab{a}})\citenamefont {Arnold}, \citenamefont {Moore},\
  and\ \citenamefont {Yaffe}}]{Arnold:2002zm}%
  \BibitemOpen
  \bibfield  {author} {\bibinfo {author} {\bibfnamefont {P.~B.}\ \bibnamefont
  {Arnold}}, \bibinfo {author} {\bibfnamefont {G.~D.}\ \bibnamefont {Moore}},\
  and\ \bibinfo {author} {\bibfnamefont {L.~G.}\ \bibnamefont {Yaffe}},\
  }\bibfield  {title} {\bibinfo {title} {{Effective kinetic theory for high
  temperature gauge theories}},\ }\href
  {https://doi.org/10.1088/1126-6708/2003/01/030} {\bibfield  {journal}
  {\bibinfo  {journal} {JHEP}\ }\textbf {\bibinfo {volume} {01}},\ \bibinfo
  {pages} {030}},\ \Eprint {https://arxiv.org/abs/hep-ph/0209353}
  {arXiv:hep-ph/0209353} \BibitemShut {NoStop}%
\bibitem [{\citenamefont {Kurkela}\ and\ \citenamefont
  {Mazeliauskas}(2019)}]{Kurkela:2018oqw}%
  \BibitemOpen
  \bibfield  {author} {\bibinfo {author} {\bibfnamefont {A.}~\bibnamefont
  {Kurkela}}\ and\ \bibinfo {author} {\bibfnamefont {A.}~\bibnamefont
  {Mazeliauskas}},\ }\bibfield  {title} {\bibinfo {title} {{Chemical
  equilibration in weakly coupled QCD}},\ }\href
  {https://doi.org/10.1103/PhysRevD.99.054018} {\bibfield  {journal} {\bibinfo
  {journal} {Phys. Rev. D}\ }\textbf {\bibinfo {volume} {99}},\ \bibinfo
  {pages} {054018} (\bibinfo {year} {2019})},\ \Eprint
  {https://arxiv.org/abs/1811.03068} {arXiv:1811.03068 [hep-ph]} \BibitemShut
  {NoStop}%
\bibitem [{\citenamefont {Landau}\ and\ \citenamefont
  {Pomeranchuk}(1953{\natexlab{a}})}]{Landau:1953gr}%
  \BibitemOpen
  \bibfield  {author} {\bibinfo {author} {\bibfnamefont {L.}~\bibnamefont
  {Landau}}\ and\ \bibinfo {author} {\bibfnamefont {I.}~\bibnamefont
  {Pomeranchuk}},\ }\bibfield  {title} {\bibinfo {title} {{Electron cascade
  process at very high-energies}},\ }\href@noop {} {\bibfield  {journal}
  {\bibinfo  {journal} {Dokl. Akad. Nauk Ser. Fiz.}\ }\textbf {\bibinfo
  {volume} {92}},\ \bibinfo {pages} {735} (\bibinfo {year}
  {1953}{\natexlab{a}})}\BibitemShut {NoStop}%
\bibitem [{\citenamefont {Landau}\ and\ \citenamefont
  {Pomeranchuk}(1953{\natexlab{b}})}]{Landau:1953um}%
  \BibitemOpen
  \bibfield  {author} {\bibinfo {author} {\bibfnamefont {L.}~\bibnamefont
  {Landau}}\ and\ \bibinfo {author} {\bibfnamefont {I.}~\bibnamefont
  {Pomeranchuk}},\ }\bibfield  {title} {\bibinfo {title} {{Limits of
  applicability of the theory of bremsstrahlung electrons and pair production
  at high-energies}},\ }\href@noop {} {\bibfield  {journal} {\bibinfo
  {journal} {Dokl. Akad. Nauk Ser. Fiz.}\ }\textbf {\bibinfo {volume} {92}},\
  \bibinfo {pages} {535} (\bibinfo {year} {1953}{\natexlab{b}})}\BibitemShut
  {NoStop}%
\bibitem [{\citenamefont {Migdal}(1955)}]{Migdal:1955nv}%
  \BibitemOpen
  \bibfield  {author} {\bibinfo {author} {\bibfnamefont {A.~B.}\ \bibnamefont
  {Migdal}},\ }\bibfield  {title} {\bibinfo {title} {{Quantum kinetic equation
  for multiple scattering}},\ }\href@noop {} {\bibfield  {journal} {\bibinfo
  {journal} {Dokl. Akad. Nauk Ser. Fiz.}\ }\textbf {\bibinfo {volume} {105}},\
  \bibinfo {pages} {77} (\bibinfo {year} {1955})}\BibitemShut {NoStop}%
\bibitem [{\citenamefont {Du}\ and\ \citenamefont
  {Schlichting}(2021{\natexlab{b}})}]{Du:2020dvp}%
  \BibitemOpen
  \bibfield  {author} {\bibinfo {author} {\bibfnamefont {X.}~\bibnamefont
  {Du}}\ and\ \bibinfo {author} {\bibfnamefont {S.}~\bibnamefont
  {Schlichting}},\ }\bibfield  {title} {\bibinfo {title} {{Equilibration of
  weakly coupled QCD plasmas}},\ }\href
  {https://doi.org/10.1103/PhysRevD.104.054011} {\bibfield  {journal} {\bibinfo
   {journal} {Phys. Rev. D}\ }\textbf {\bibinfo {volume} {104}},\ \bibinfo
  {pages} {054011} (\bibinfo {year} {2021}{\natexlab{b}})},\ \Eprint
  {https://arxiv.org/abs/2012.09079} {arXiv:2012.09079 [hep-ph]} \BibitemShut
  {NoStop}%
\bibitem [{\citenamefont {Abraao~York}\ \emph {et~al.}(2014)\citenamefont
  {Abraao~York}, \citenamefont {Kurkela}, \citenamefont {Lu},\ and\
  \citenamefont {Moore}}]{AbraaoYork:2014hbk}%
  \BibitemOpen
  \bibfield  {author} {\bibinfo {author} {\bibfnamefont {M.~C.}\ \bibnamefont
  {Abraao~York}}, \bibinfo {author} {\bibfnamefont {A.}~\bibnamefont
  {Kurkela}}, \bibinfo {author} {\bibfnamefont {E.}~\bibnamefont {Lu}},\ and\
  \bibinfo {author} {\bibfnamefont {G.~D.}\ \bibnamefont {Moore}},\ }\bibfield
  {title} {\bibinfo {title} {{UV cascade in classical Yang-Mills theory via
  kinetic theory}},\ }\href {https://doi.org/10.1103/PhysRevD.89.074036}
  {\bibfield  {journal} {\bibinfo  {journal} {Phys. Rev. D}\ }\textbf {\bibinfo
  {volume} {89}},\ \bibinfo {pages} {074036} (\bibinfo {year} {2014})},\
  \Eprint {https://arxiv.org/abs/1401.3751} {arXiv:1401.3751 [hep-ph]}
  \BibitemShut {NoStop}%
\bibitem [{\citenamefont {Heller}\ \emph {et~al.}(2012)\citenamefont {Heller},
  \citenamefont {Janik},\ and\ \citenamefont {Witaszczyk}}]{Heller:2011ju}%
  \BibitemOpen
  \bibfield  {author} {\bibinfo {author} {\bibfnamefont {M.~P.}\ \bibnamefont
  {Heller}}, \bibinfo {author} {\bibfnamefont {R.~A.}\ \bibnamefont {Janik}},\
  and\ \bibinfo {author} {\bibfnamefont {P.}~\bibnamefont {Witaszczyk}},\
  }\bibfield  {title} {\bibinfo {title} {{The characteristics of thermalization
  of boost-invariant plasma from holography}},\ }\href
  {https://doi.org/10.1103/PhysRevLett.108.201602} {\bibfield  {journal}
  {\bibinfo  {journal} {Phys. Rev. Lett.}\ }\textbf {\bibinfo {volume} {108}},\
  \bibinfo {pages} {201602} (\bibinfo {year} {2012})},\ \Eprint
  {https://arxiv.org/abs/1103.3452} {arXiv:1103.3452 [hep-th]} \BibitemShut
  {NoStop}%
\bibitem [{\citenamefont {Keegan}\ \emph {et~al.}(2016)\citenamefont {Keegan},
  \citenamefont {Kurkela}, \citenamefont {Romatschke}, \citenamefont {van~der
  Schee},\ and\ \citenamefont {Zhu}}]{Keegan:2015avk}%
  \BibitemOpen
  \bibfield  {author} {\bibinfo {author} {\bibfnamefont {L.}~\bibnamefont
  {Keegan}}, \bibinfo {author} {\bibfnamefont {A.}~\bibnamefont {Kurkela}},
  \bibinfo {author} {\bibfnamefont {P.}~\bibnamefont {Romatschke}}, \bibinfo
  {author} {\bibfnamefont {W.}~\bibnamefont {van~der Schee}},\ and\ \bibinfo
  {author} {\bibfnamefont {Y.}~\bibnamefont {Zhu}},\ }\bibfield  {title}
  {\bibinfo {title} {{Weak and strong coupling equilibration in nonabelian
  gauge theories}},\ }\href {https://doi.org/10.1007/JHEP04(2016)031}
  {\bibfield  {journal} {\bibinfo  {journal} {JHEP}\ }\textbf {\bibinfo
  {volume} {04}},\ \bibinfo {pages} {031}},\ \Eprint
  {https://arxiv.org/abs/1512.05347} {arXiv:1512.05347 [hep-th]} \BibitemShut
  {NoStop}%
\bibitem [{\citenamefont {Heller}\ \emph
  {et~al.}(2018{\natexlab{b}})\citenamefont {Heller}, \citenamefont {Kurkela},
  \citenamefont {Spali\'nski},\ and\ \citenamefont
  {Svensson}}]{Heller:2016rtz}%
  \BibitemOpen
  \bibfield  {author} {\bibinfo {author} {\bibfnamefont {M.~P.}\ \bibnamefont
  {Heller}}, \bibinfo {author} {\bibfnamefont {A.}~\bibnamefont {Kurkela}},
  \bibinfo {author} {\bibfnamefont {M.}~\bibnamefont {Spali\'nski}},\ and\
  \bibinfo {author} {\bibfnamefont {V.}~\bibnamefont {Svensson}},\ }\bibfield
  {title} {\bibinfo {title} {{Hydrodynamization in kinetic theory: Transient
  modes and the gradient expansion}},\ }\href
  {https://doi.org/10.1103/PhysRevD.97.091503} {\bibfield  {journal} {\bibinfo
  {journal} {Phys. Rev. D}\ }\textbf {\bibinfo {volume} {97}},\ \bibinfo
  {pages} {091503} (\bibinfo {year} {2018}{\natexlab{b}})},\ \Eprint
  {https://arxiv.org/abs/1609.04803} {arXiv:1609.04803 [nucl-th]} \BibitemShut
  {NoStop}%
\bibitem [{\citenamefont {Policastro}\ \emph {et~al.}(2001)\citenamefont
  {Policastro}, \citenamefont {Son},\ and\ \citenamefont
  {Starinets}}]{Policastro:2001yc}%
  \BibitemOpen
  \bibfield  {author} {\bibinfo {author} {\bibfnamefont {G.}~\bibnamefont
  {Policastro}}, \bibinfo {author} {\bibfnamefont {D.~T.}\ \bibnamefont
  {Son}},\ and\ \bibinfo {author} {\bibfnamefont {A.~O.}\ \bibnamefont
  {Starinets}},\ }\bibfield  {title} {\bibinfo {title} {{The Shear viscosity of
  strongly coupled N=4 supersymmetric Yang-Mills plasma}},\ }\href
  {https://doi.org/10.1103/PhysRevLett.87.081601} {\bibfield  {journal}
  {\bibinfo  {journal} {Phys. Rev. Lett.}\ }\textbf {\bibinfo {volume} {87}},\
  \bibinfo {pages} {081601} (\bibinfo {year} {2001})},\ \Eprint
  {https://arxiv.org/abs/hep-th/0104066} {arXiv:hep-th/0104066} \BibitemShut
  {NoStop}%
\bibitem [{\citenamefont {Kovtun}\ \emph {et~al.}(2005)\citenamefont {Kovtun},
  \citenamefont {Son},\ and\ \citenamefont {Starinets}}]{Kovtun:2004de}%
  \BibitemOpen
  \bibfield  {author} {\bibinfo {author} {\bibfnamefont {P.}~\bibnamefont
  {Kovtun}}, \bibinfo {author} {\bibfnamefont {D.~T.}\ \bibnamefont {Son}},\
  and\ \bibinfo {author} {\bibfnamefont {A.~O.}\ \bibnamefont {Starinets}},\
  }\bibfield  {title} {\bibinfo {title} {{Viscosity in strongly interacting
  quantum field theories from black hole physics}},\ }\href
  {https://doi.org/10.1103/PhysRevLett.94.111601} {\bibfield  {journal}
  {\bibinfo  {journal} {Phys. Rev. Lett.}\ }\textbf {\bibinfo {volume} {94}},\
  \bibinfo {pages} {111601} (\bibinfo {year} {2005})},\ \Eprint
  {https://arxiv.org/abs/hep-th/0405231} {arXiv:hep-th/0405231} \BibitemShut
  {NoStop}%
\bibitem [{\citenamefont {Romatschke}\ and\ \citenamefont
  {Strickland}(2003)}]{Romatschke:2003ms}%
  \BibitemOpen
  \bibfield  {author} {\bibinfo {author} {\bibfnamefont {P.}~\bibnamefont
  {Romatschke}}\ and\ \bibinfo {author} {\bibfnamefont {M.}~\bibnamefont
  {Strickland}},\ }\bibfield  {title} {\bibinfo {title} {{Collective modes of
  an anisotropic quark gluon plasma}},\ }\href
  {https://doi.org/10.1103/PhysRevD.68.036004} {\bibfield  {journal} {\bibinfo
  {journal} {Phys. Rev. D}\ }\textbf {\bibinfo {volume} {68}},\ \bibinfo
  {pages} {036004} (\bibinfo {year} {2003})},\ \Eprint
  {https://arxiv.org/abs/hep-ph/0304092} {arXiv:hep-ph/0304092} \BibitemShut
  {NoStop}%
\bibitem [{\citenamefont {Ba{\c s}ar}\ and\ \citenamefont
  {Dunne}(2015)}]{Basar:2015ava}%
  \BibitemOpen
  \bibfield  {author} {\bibinfo {author} {\bibfnamefont {G.}~\bibnamefont
  {Ba{\c s}ar}}\ and\ \bibinfo {author} {\bibfnamefont {G.~V.}\ \bibnamefont
  {Dunne}},\ }\bibfield  {title} {\bibinfo {title} {{Hydrodynamics, resurgence,
  and transasymptotics}},\ }\href {https://doi.org/10.1103/PhysRevD.92.125011}
  {\bibfield  {journal} {\bibinfo  {journal} {Phys. Rev.}\ }\textbf {\bibinfo
  {volume} {D92}},\ \bibinfo {pages} {125011} (\bibinfo {year} {2015})},\
  \Eprint {https://arxiv.org/abs/1509.05046} {arXiv:1509.05046 [hep-th]}
  \BibitemShut {NoStop}%
%%CITATION = ARXIV:1509.05046;%%
\bibitem [{\citenamefont {Aniceto}\ and\ \citenamefont
  {Spali{\'n}ski}(2016)}]{Aniceto:2015mto}%
  \BibitemOpen
  \bibfield  {author} {\bibinfo {author} {\bibfnamefont {I.}~\bibnamefont
  {Aniceto}}\ and\ \bibinfo {author} {\bibfnamefont {M.}~\bibnamefont
  {Spali{\'n}ski}},\ }\bibfield  {title} {\bibinfo {title} {{Resurgence in
  Extended Hydrodynamics}},\ }\href
  {https://doi.org/10.1103/PhysRevD.93.085008} {\bibfield  {journal} {\bibinfo
  {journal} {Phys. Rev.}\ }\textbf {\bibinfo {volume} {D93}},\ \bibinfo {pages}
  {085008} (\bibinfo {year} {2016})},\ \Eprint
  {https://arxiv.org/abs/1511.06358} {arXiv:1511.06358 [hep-th]} \BibitemShut
  {NoStop}%
%%CITATION = ARXIV:1511.06358;%%
\bibitem [{\citenamefont {Muller}(1967)}]{Muller:1967zza}%
  \BibitemOpen
  \bibfield  {author} {\bibinfo {author} {\bibfnamefont {I.}~\bibnamefont
  {Muller}},\ }\bibfield  {title} {\bibinfo {title} {{Zum Paradoxon der
  Warmeleitungstheorie}},\ }\href {https://doi.org/10.1007/BF01326412}
  {\bibfield  {journal} {\bibinfo  {journal} {Z.Phys.}\ }\textbf {\bibinfo
  {volume} {198}},\ \bibinfo {pages} {329} (\bibinfo {year}
  {1967})}\BibitemShut {NoStop}%
%%CITATION = ZEPYA,198,329;%%
\bibitem [{\citenamefont {Israel}(1976)}]{Israel1976Sep}%
  \BibitemOpen
  \bibfield  {author} {\bibinfo {author} {\bibfnamefont {W.}~\bibnamefont
  {Israel}},\ }\bibfield  {title} {\bibinfo {title} {Nonstationary irreversible
  thermodynamics: a causal relativistic theory},\ }\href@noop {} {\bibfield
  {journal} {\bibinfo  {journal} {Annals of Physics}\ }\textbf {\bibinfo
  {volume} {100}},\ \bibinfo {pages} {310} (\bibinfo {year}
  {1976})}\BibitemShut {NoStop}%
\bibitem [{\citenamefont {Israel}\ and\ \citenamefont
  {Stewart}(1979)}]{Israel:1979wp}%
  \BibitemOpen
  \bibfield  {author} {\bibinfo {author} {\bibfnamefont {W.}~\bibnamefont
  {Israel}}\ and\ \bibinfo {author} {\bibfnamefont {J.}~\bibnamefont
  {Stewart}},\ }\bibfield  {title} {\bibinfo {title} {{Transient relativistic
  thermodynamics and kinetic theory}},\ }\href
  {https://doi.org/10.1016/0003-4916(79)90130-1} {\bibfield  {journal}
  {\bibinfo  {journal} {Annals Phys.}\ }\textbf {\bibinfo {volume} {118}},\
  \bibinfo {pages} {341} (\bibinfo {year} {1979})}\BibitemShut {NoStop}%
%%CITATION = APNYA,118,341;%%
\bibitem [{\citenamefont {Aniceto}\ \emph
  {et~al.}(2019{\natexlab{a}})\citenamefont {Aniceto}, \citenamefont {Basar},\
  and\ \citenamefont {Schiappa}}]{Aniceto:2018bis}%
  \BibitemOpen
  \bibfield  {author} {\bibinfo {author} {\bibfnamefont {I.}~\bibnamefont
  {Aniceto}}, \bibinfo {author} {\bibfnamefont {G.}~\bibnamefont {Basar}},\
  and\ \bibinfo {author} {\bibfnamefont {R.}~\bibnamefont {Schiappa}},\
  }\bibfield  {title} {\bibinfo {title} {{A Primer on Resurgent Transseries and
  Their Asymptotics}},\ }\href {https://doi.org/10.1016/j.physrep.2019.02.003}
  {\bibfield  {journal} {\bibinfo  {journal} {Phys. Rept.}\ }\textbf {\bibinfo
  {volume} {809}},\ \bibinfo {pages} {1} (\bibinfo {year}
  {2019}{\natexlab{a}})},\ \Eprint {https://arxiv.org/abs/1802.10441}
  {arXiv:1802.10441 [hep-th]} \BibitemShut {NoStop}%
\bibitem [{\citenamefont {Heller}\ \emph {et~al.}(2021)\citenamefont {Heller},
  \citenamefont {Serantes}, \citenamefont {Spali\'nski}, \citenamefont
  {Svensson},\ and\ \citenamefont {Withers}}]{Heller:2021yjh}%
  \BibitemOpen
  \bibfield  {author} {\bibinfo {author} {\bibfnamefont {M.~P.}\ \bibnamefont
  {Heller}}, \bibinfo {author} {\bibfnamefont {A.}~\bibnamefont {Serantes}},
  \bibinfo {author} {\bibfnamefont {M.}~\bibnamefont {Spali\'nski}}, \bibinfo
  {author} {\bibfnamefont {V.}~\bibnamefont {Svensson}},\ and\ \bibinfo
  {author} {\bibfnamefont {B.}~\bibnamefont {Withers}},\ }\bibfield  {title}
  {\bibinfo {title} {{Relativistic hydrodynamics: a singulant perspective}},\
  }\href@noop {} {\  (\bibinfo {year} {2021})},\ \Eprint
  {https://arxiv.org/abs/2112.12794} {arXiv:2112.12794 [hep-th]} \BibitemShut
  {NoStop}%
\bibitem [{\citenamefont {Heller}\ \emph {et~al.}(2013)\citenamefont {Heller},
  \citenamefont {Janik},\ and\ \citenamefont {Witaszczyk}}]{Heller:2013fn}%
  \BibitemOpen
  \bibfield  {author} {\bibinfo {author} {\bibfnamefont {M.~P.}\ \bibnamefont
  {Heller}}, \bibinfo {author} {\bibfnamefont {R.~A.}\ \bibnamefont {Janik}},\
  and\ \bibinfo {author} {\bibfnamefont {P.}~\bibnamefont {Witaszczyk}},\
  }\bibfield  {title} {\bibinfo {title} {{Hydrodynamic Gradient Expansion in
  Gauge Theory Plasmas}},\ }\href
  {https://doi.org/10.1103/PhysRevLett.110.211602} {\bibfield  {journal}
  {\bibinfo  {journal} {Phys. Rev. Lett.}\ }\textbf {\bibinfo {volume} {110}},\
  \bibinfo {pages} {211602} (\bibinfo {year} {2013})},\ \Eprint
  {https://arxiv.org/abs/1302.0697} {arXiv:1302.0697 [hep-th]} \BibitemShut
  {NoStop}%
\bibitem [{\citenamefont {{Casalderrey-Solana}}\ \emph
  {et~al.}(2018)\citenamefont {{Casalderrey-Solana}}, \citenamefont
  {Gushterov},\ and\ \citenamefont
  {Meiring}}]{casalderreysolanaResurgenceHydrodynamicAttractors2018}%
  \BibitemOpen
  \bibfield  {author} {\bibinfo {author} {\bibfnamefont {J.}~\bibnamefont
  {{Casalderrey-Solana}}}, \bibinfo {author} {\bibfnamefont {N.~I.}\
  \bibnamefont {Gushterov}},\ and\ \bibinfo {author} {\bibfnamefont
  {B.}~\bibnamefont {Meiring}},\ }\bibfield  {title} {\bibinfo {title}
  {Resurgence and hydrodynamic attractors in {{Gauss-Bonnet}} holography},\
  }\href {https://doi.org/10.1007/JHEP04(2018)042} {\bibfield  {journal}
  {\bibinfo  {journal} {Journal of High Energy Physics}\ }\textbf {\bibinfo
  {volume} {2018}},\ \bibinfo {pages} {42} (\bibinfo {year}
  {2018})}\BibitemShut {NoStop}%
\bibitem [{\citenamefont {Aniceto}\ \emph
  {et~al.}(2019{\natexlab{b}})\citenamefont {Aniceto}, \citenamefont
  {Jankowski}, \citenamefont {Meiring},\ and\ \citenamefont
  {Spali{\'n}ski}}]{anicetoLargePropertimeExpansion2019}%
  \BibitemOpen
  \bibfield  {author} {\bibinfo {author} {\bibfnamefont {I.}~\bibnamefont
  {Aniceto}}, \bibinfo {author} {\bibfnamefont {J.}~\bibnamefont {Jankowski}},
  \bibinfo {author} {\bibfnamefont {B.}~\bibnamefont {Meiring}},\ and\ \bibinfo
  {author} {\bibfnamefont {M.}~\bibnamefont {Spali{\'n}ski}},\ }\bibfield
  {title} {\bibinfo {title} {The large proper-time expansion of {{Yang-Mills}}
  plasma as a resurgent transseries},\ }\href
  {https://doi.org/10.1007/JHEP02(2019)073} {\bibfield  {journal} {\bibinfo
  {journal} {Journal of High Energy Physics}\ }\textbf {\bibinfo {volume}
  {2019}},\ \bibinfo {pages} {73} (\bibinfo {year}
  {2019}{\natexlab{b}})}\BibitemShut {NoStop}%
\bibitem [{\citenamefont {Kovtun}\ and\ \citenamefont
  {Starinets}(2005)}]{Kovtun:2005ev}%
  \BibitemOpen
  \bibfield  {author} {\bibinfo {author} {\bibfnamefont {P.~K.}\ \bibnamefont
  {Kovtun}}\ and\ \bibinfo {author} {\bibfnamefont {A.~O.}\ \bibnamefont
  {Starinets}},\ }\bibfield  {title} {\bibinfo {title} {{Quasinormal modes and
  holography}},\ }\href {https://doi.org/10.1103/PhysRevD.72.086009} {\bibfield
   {journal} {\bibinfo  {journal} {Phys. Rev. D}\ }\textbf {\bibinfo {volume}
  {72}},\ \bibinfo {pages} {086009} (\bibinfo {year} {2005})},\ \Eprint
  {https://arxiv.org/abs/hep-th/0506184} {arXiv:hep-th/0506184} \BibitemShut
  {NoStop}%
\bibitem [{\citenamefont {Janik}\ and\ \citenamefont
  {Peschanski}(2006)}]{Janik:2006gp}%
  \BibitemOpen
  \bibfield  {author} {\bibinfo {author} {\bibfnamefont {R.~A.}\ \bibnamefont
  {Janik}}\ and\ \bibinfo {author} {\bibfnamefont {R.~B.}\ \bibnamefont
  {Peschanski}},\ }\bibfield  {title} {\bibinfo {title} {{Gauge/gravity duality
  and thermalization of a boost-invariant perfect fluid}},\ }\href
  {https://doi.org/10.1103/PhysRevD.74.046007} {\bibfield  {journal} {\bibinfo
  {journal} {Phys. Rev. D}\ }\textbf {\bibinfo {volume} {74}},\ \bibinfo
  {pages} {046007} (\bibinfo {year} {2006})},\ \Eprint
  {https://arxiv.org/abs/hep-th/0606149} {arXiv:hep-th/0606149} \BibitemShut
  {NoStop}%
\bibitem [{\citenamefont {Kurkela}\ and\ \citenamefont
  {Wiedemann}(2019)}]{Kurkela:2017xis}%
  \BibitemOpen
  \bibfield  {author} {\bibinfo {author} {\bibfnamefont {A.}~\bibnamefont
  {Kurkela}}\ and\ \bibinfo {author} {\bibfnamefont {U.~A.}\ \bibnamefont
  {Wiedemann}},\ }\bibfield  {title} {\bibinfo {title} {{Analytic structure of
  nonhydrodynamic modes in kinetic theory}},\ }\href
  {https://doi.org/10.1140/epjc/s10052-019-7271-9} {\bibfield  {journal}
  {\bibinfo  {journal} {Eur. Phys. J. C}\ }\textbf {\bibinfo {volume} {79}},\
  \bibinfo {pages} {776} (\bibinfo {year} {2019})},\ \Eprint
  {https://arxiv.org/abs/1712.04376} {arXiv:1712.04376 [hep-ph]} \BibitemShut
  {NoStop}%
\bibitem [{\citenamefont {Moore}(2018)}]{Moore:2018mma}%
  \BibitemOpen
  \bibfield  {author} {\bibinfo {author} {\bibfnamefont {G.~D.}\ \bibnamefont
  {Moore}},\ }\bibfield  {title} {\bibinfo {title} {{Stress-stress correlator
  in $\phi^{4}$ theory: poles or a cut?}},\ }\href
  {https://doi.org/10.1007/JHEP05(2018)084} {\bibfield  {journal} {\bibinfo
  {journal} {JHEP}\ }\textbf {\bibinfo {volume} {05}},\ \bibinfo {pages}
  {084}},\ \Eprint {https://arxiv.org/abs/1803.00736} {arXiv:1803.00736
  [hep-ph]} \BibitemShut {NoStop}%
\bibitem [{\citenamefont {York}\ and\ \citenamefont
  {Moore}(2009)}]{York:2008rr}%
  \BibitemOpen
  \bibfield  {author} {\bibinfo {author} {\bibfnamefont {M.~A.}\ \bibnamefont
  {York}}\ and\ \bibinfo {author} {\bibfnamefont {G.~D.}\ \bibnamefont
  {Moore}},\ }\bibfield  {title} {\bibinfo {title} {{Second order hydrodynamic
  coefficients from kinetic theory}},\ }\href
  {https://doi.org/10.1103/PhysRevD.79.054011} {\bibfield  {journal} {\bibinfo
  {journal} {Phys. Rev. D}\ }\textbf {\bibinfo {volume} {79}},\ \bibinfo
  {pages} {054011} (\bibinfo {year} {2009})},\ \Eprint
  {https://arxiv.org/abs/0811.0729} {arXiv:0811.0729 [hep-ph]} \BibitemShut
  {NoStop}%
\bibitem [{\citenamefont {Kamata}\ \emph {et~al.}(2020)\citenamefont {Kamata},
  \citenamefont {Martinez}, \citenamefont {Plaschke}, \citenamefont
  {Ochsenfeld},\ and\ \citenamefont {Schlichting}}]{Kamata:2020mka}%
  \BibitemOpen
  \bibfield  {author} {\bibinfo {author} {\bibfnamefont {S.}~\bibnamefont
  {Kamata}}, \bibinfo {author} {\bibfnamefont {M.}~\bibnamefont {Martinez}},
  \bibinfo {author} {\bibfnamefont {P.}~\bibnamefont {Plaschke}}, \bibinfo
  {author} {\bibfnamefont {S.}~\bibnamefont {Ochsenfeld}},\ and\ \bibinfo
  {author} {\bibfnamefont {S.}~\bibnamefont {Schlichting}},\ }\bibfield
  {title} {\bibinfo {title} {{Hydrodynamization and nonequilibrium
  Green\textquoteright{}s functions in kinetic theory}},\ }\href
  {https://doi.org/10.1103/PhysRevD.102.056003} {\bibfield  {journal} {\bibinfo
   {journal} {Phys. Rev. D}\ }\textbf {\bibinfo {volume} {102}},\ \bibinfo
  {pages} {056003} (\bibinfo {year} {2020})},\ \Eprint
  {https://arxiv.org/abs/2004.06751} {arXiv:2004.06751 [hep-ph]} \BibitemShut
  {NoStop}%
\bibitem [{\citenamefont {Kurkela}\ \emph
  {et~al.}(2019{\natexlab{c}})\citenamefont {Kurkela}, \citenamefont
  {Wiedemann},\ and\ \citenamefont {Wu}}]{Kurkela:2018qeb}%
  \BibitemOpen
  \bibfield  {author} {\bibinfo {author} {\bibfnamefont {A.}~\bibnamefont
  {Kurkela}}, \bibinfo {author} {\bibfnamefont {U.~A.}\ \bibnamefont
  {Wiedemann}},\ and\ \bibinfo {author} {\bibfnamefont {B.}~\bibnamefont
  {Wu}},\ }\bibfield  {title} {\bibinfo {title} {{Opacity dependence of
  elliptic flow in kinetic theory}},\ }\href
  {https://doi.org/10.1140/epjc/s10052-019-7262-x} {\bibfield  {journal}
  {\bibinfo  {journal} {Eur. Phys. J. C}\ }\textbf {\bibinfo {volume} {79}},\
  \bibinfo {pages} {759} (\bibinfo {year} {2019}{\natexlab{c}})},\ \Eprint
  {https://arxiv.org/abs/1805.04081} {arXiv:1805.04081 [hep-ph]} \BibitemShut
  {NoStop}%
\bibitem [{\citenamefont {Ambrus}\ \emph {et~al.}(2022)\citenamefont {Ambrus},
  \citenamefont {Schlichting},\ and\ \citenamefont
  {Werthmann}}]{Ambrus:2021fej}%
  \BibitemOpen
  \bibfield  {author} {\bibinfo {author} {\bibfnamefont {V.~E.}\ \bibnamefont
  {Ambrus}}, \bibinfo {author} {\bibfnamefont {S.}~\bibnamefont
  {Schlichting}},\ and\ \bibinfo {author} {\bibfnamefont {C.}~\bibnamefont
  {Werthmann}},\ }\bibfield  {title} {\bibinfo {title} {{Development of
  transverse flow at small and large opacities in conformal kinetic theory}},\
  }\href {https://doi.org/10.1103/PhysRevD.105.014031} {\bibfield  {journal}
  {\bibinfo  {journal} {Phys. Rev. D}\ }\textbf {\bibinfo {volume} {105}},\
  \bibinfo {pages} {014031} (\bibinfo {year} {2022})},\ \Eprint
  {https://arxiv.org/abs/2109.03290} {arXiv:2109.03290 [hep-ph]} \BibitemShut
  {NoStop}%
\bibitem [{\citenamefont {Romatschke}(2016)}]{Romatschke:2015gic}%
  \BibitemOpen
  \bibfield  {author} {\bibinfo {author} {\bibfnamefont {P.}~\bibnamefont
  {Romatschke}},\ }\bibfield  {title} {\bibinfo {title} {{Retarded correlators
  in kinetic theory: branch cuts, poles and hydrodynamic onset transitions}},\
  }\href {https://doi.org/10.1140/epjc/s10052-016-4169-7} {\bibfield  {journal}
  {\bibinfo  {journal} {Eur. Phys. J. C}\ }\textbf {\bibinfo {volume} {76}},\
  \bibinfo {pages} {352} (\bibinfo {year} {2016})},\ \Eprint
  {https://arxiv.org/abs/1512.02641} {arXiv:1512.02641 [hep-th]} \BibitemShut
  {NoStop}%
\bibitem [{\citenamefont {Rocha}\ \emph {et~al.}(2021)\citenamefont {Rocha},
  \citenamefont {Denicol},\ and\ \citenamefont {Noronha}}]{Rocha:2021zcw}%
  \BibitemOpen
  \bibfield  {author} {\bibinfo {author} {\bibfnamefont {G.~S.}\ \bibnamefont
  {Rocha}}, \bibinfo {author} {\bibfnamefont {G.~S.}\ \bibnamefont {Denicol}},\
  and\ \bibinfo {author} {\bibfnamefont {J.}~\bibnamefont {Noronha}},\
  }\bibfield  {title} {\bibinfo {title} {{Novel Relaxation Time Approximation
  to the Relativistic Boltzmann Equation}},\ }\href
  {https://doi.org/10.1103/PhysRevLett.127.042301} {\bibfield  {journal}
  {\bibinfo  {journal} {Phys. Rev. Lett.}\ }\textbf {\bibinfo {volume} {127}},\
  \bibinfo {pages} {042301} (\bibinfo {year} {2021})},\ \Eprint
  {https://arxiv.org/abs/2103.07489} {arXiv:2103.07489 [nucl-th]} \BibitemShut
  {NoStop}%
\bibitem [{\citenamefont {Arnold}\ \emph
  {et~al.}(2003{\natexlab{b}})\citenamefont {Arnold}, \citenamefont {Moore},\
  and\ \citenamefont {Yaffe}}]{Arnold:2003zc}%
  \BibitemOpen
  \bibfield  {author} {\bibinfo {author} {\bibfnamefont {P.~B.}\ \bibnamefont
  {Arnold}}, \bibinfo {author} {\bibfnamefont {G.~D.}\ \bibnamefont {Moore}},\
  and\ \bibinfo {author} {\bibfnamefont {L.~G.}\ \bibnamefont {Yaffe}},\
  }\bibfield  {title} {\bibinfo {title} {{Transport coefficients in high
  temperature gauge theories. 2. Beyond leading log}},\ }\href
  {https://doi.org/10.1088/1126-6708/2003/05/051} {\bibfield  {journal}
  {\bibinfo  {journal} {JHEP}\ }\textbf {\bibinfo {volume} {05}},\ \bibinfo
  {pages} {051}},\ \Eprint {https://arxiv.org/abs/hep-ph/0302165}
  {arXiv:hep-ph/0302165} \BibitemShut {NoStop}%
\end{thebibliography}%
\end{document}